\newif\ifextended
\extendedtrue

 \documentclass[acmsmall,screen]{acmart}

\setcopyright{rightsretained}
\acmPrice{}
\acmDOI{10.1145/3527327}
\acmYear{2022}
\copyrightyear{2022}
\acmSubmissionID{oopsla22main-p90-p}
\acmJournal{PACMPL}
\acmVolume{6}
\acmNumber{OOPSLA1}
\acmArticle{83}
\acmMonth{4}

\usepackage{booktabs}   \usepackage{subcaption}

\usepackage{microtype}
\usepackage{setspace}
\usepackage{xspace}
\usepackage{listings}
\usepackage{lstcoq}
\usepackage{lstjavascript}

\usepackage{parcolumns}
\usepackage{calc}
\usepackage{amsmath}
\usepackage{multirow}
\usepackage{multicol}
\usepackage{proof}
\usepackage{calrsfs}
\usepackage{pifont}
\usepackage{mathtools}
\usepackage[inline]{enumitem}
\usepackage{stmaryrd}

\usepackage{tikz}
\usepackage{calc}
\usetikzlibrary{automata}
\usetikzlibrary{positioning,calc,decorations.markings,shapes,graphs,
  snakes,arrows}
\usepackage[nosort]{cleveref}
\usepackage{hyperref}

\renewcommand{\marginpar}[1]{}

\definecolor{darkblue}{rgb}{0,0.2,0.4}

\definecolor{denim}{rgb}{0.08, 0.38, 0.74}
 
\definecolor{darkred}{rgb}{0.4,0,0.2}

\newcommand{\eg}{\emph{e.g.}, }
\newcommand{\ie}{\emph{i.e.}, }

\newcommand{\wrt}{\emph{w.r.t} }

\newcommand{\coq}{Coq\xspace}
\newcommand{\js}{JavaScript\xspace}
\newcommand{\sql}{\mbox{SQL}\xspace}
\newcommand{\sqlalg}{SQL$_{\mbox{\tiny Alg}}$\xspace}
\newcommand{\sqlcoq}{SQL$_{\mbox{\tiny Coq}}$\xspace}
\newcommand{\nnrc}{NNRC\xspace}
\newcommand{\imp}{Imp\xspace}

\renewcommand{\null}{\protect{\tt NULL}\xspace}

\newcommand{\attribute}{a}

\newcommand{\formula}{f}

\newcommand{\ag}{\ensuremath{\mathfrak{ag}}}
\newcommand{\fn}{\ensuremath{\mathfrak{fn}}}
\newcommand{\pr}{\ensuremath{\mathfrak{pr}}}
\newcommand{\ea}{\ensuremath{e^a}}
\newcommand{\ef}{\ensuremath{e^f}}

\newcommand{\env}{\ensuremath{\mathcal{E}}}

\newcommand{\isbuiltupon}[2]{\ensuremath{{\mathbb B}_{u}(#1,#2)}}
\newcommand{\values}{\ensuremath{\mathcal V}}

\newcommand{\findevalenv}[2]{\ensuremath{{\mathbb F}_{e}(#1,#2)}}
\newcommand{\semf}[3]{\ensuremath{\llbracket#1\rrbracket^{\mathsf{f}}_{#2}({#3})}}
\newcommand{\sema}[2]{\ensuremath{\llbracket#1\rrbracket^{\mathsf{a}}_{#2}}}
\newcommand{\semb}[3]{\ensuremath{\llbracket#1\rrbracket^{\mathsf{b}}_{#2}({#3})}}
\newcommand{\semq}[3]{\ensuremath{\llbracket#1\rrbracket^{\mathsf{Q}}_{#2}(#3)}}
\newcommand{\seme}[2]{\ensuremath{\llbracket#1\rrbracket^{\mathsf{e}}_{#2}}}

\newcommand{\algquery}{\ensuremath{Q}}

\newcommand{\NRAEnv}{\texorpdfstring{NRA$^{\!\mbox{\it e}}$}{NRAe}\xspace}
\newcommand{\nrae}{\NRAEnv}

\usepackage{listings}

\newif\ifelec

\electrue
\definecolor{burntorange}{rgb}{0.8, 0.33, 0.0}

\lstdefinelanguage{SSR}{
mathescape=true,
texcl=false,
morekeywords=[1]{
Section, Module, End, Require, Import, Export, Defensive, Function, Axioms,
Variable, Variables, Parameter, Parameters, Axiom, Hypothesis, Hypotheses,
Notation, Local, Tactic, Reserved, Scope, Open, Close, Bind, Delimit,
Definition, Let, Ltac, Fixpoint, CoFixpoint, Add, Morphism, Relation,
Implicit, Identity, Types, Arguments, Unset, Contextual, Strict, Prenex, Implicits,
Inductive, CoInductive, Record, Structure, Canonical, Coercion,
Theorem, Lemma, Corollary, Proposition, Fact, Remark, Example,
Proof, Goal, Save, Qed, Defined, Hint, Resolve, Rewrite, View,
Search, Show, Print, Printing, All, Graph, Projections, inside,
outside, Locate, Maximal, Eval, Compute, Time, Check, Print, About,
Inline, Class, Instance, Context, Include, Declare},
morekeywords=[2]{forall, fun, fix, cofix, struct,
      match, with, end, as, in, return, let, if, is, then, else,
      for, of, nosimpl, where, True, False, beta, delta, zeta, iota},
morekeywords=[3]{Set, Type, Prop},
morekeywords=[4]{
         exists, exists2,
         pose, set, move, case, elim, apply, clear,
            hnf, intro, intros, generalize, rename, pattern, after,
	    destruct, induction, using, refine, inversion, injection,
            constructor,
         rewrite, congr, unlock, compute, vm_compute, native_compute,
            replace, fold, unfold, change, cutrewrite, simpl,
            cbv, lazy,
         have, suff, wlog, suffices, without, loss, nat_norm,
            assert, cut, trivial, revert, bool_congr, nat_congr,
	 symmetry, transitivity, auto, split, left, right,
         autorewrite,
       interval_intro},
morekeywords=[5]{
         by, done, exact, reflexivity, tauto, romega, omega,
         assumption, solve, contradiction, discriminate,
         ring, field, interval},
morekeywords=[6]{do, last, first, try, idtac, repeat, progress},
literate=
        {:=}{{$\mathrel{\mathop:\mathopen=}$}}2
{=}{{$=$}}1
{==}{{$\equiv$}}1
        {!=}{{$\not\equiv$}}1
        {<=}{{$\leq$}}1
        {>=}{{$\geq$}}1
        {<>}{{$\neq$}}1
        {->}{{$\rightarrow$}}2
{<-}{{$\leftarrow$}}2
{=>}{{$\Rightarrow$}}1
{/\\}{{$\wedge$}}2
        {\\/}{{$\vee$}}2
        {<->}{{$\leftrightarrow$}}2
        {<=>}{{$\Leftrightarrow$}}2
{forall\ }{{$\forall\,$}}1
        {exists\ }{{$\exists\,$}}1
        {negb}{{$\neg$}}1
        {~}{{$\neg$}}1
        {\\in}{{$\in$}}1
{^-1}{{$^{-1}$}}1,
comment=[s]{(*}{*)},
showstringspaces=false,
morestring=[b]",
morestring=[d]´,
extendedchars=true,
sensitive=true,
breaklines=true,
basicstyle=\footnotesize\ttfamily,
captionpos=b,
columns=[l]fullflexible,
keepspaces=true,
identifierstyle={\ttfamily\ifelec\color{black}\fi},
keywordstyle=[1]{\ttfamily\ifelec\color{dkviolet}\fi},
keywordstyle=[2]{\ttfamily\ifelec\color{dkgreen}\fi},
keywordstyle=[3]{\ttfamily\ifelec\color{dkgreen}\fi},
keywordstyle=[4]{\ttfamily\ifelec\color{dkblue}\fi},
keywordstyle=[5]{\ttfamily\ifelec\color{red}\fi},
keywordstyle=[6]{\ttfamily\ifelec\color{dkpink}\fi},
stringstyle=\ttfamily,
commentstyle={\ttfamily\ifelec\color{firebrick}\fi},
moredelim=[is][\color{burntorange}\bfseries\ttfamily]{|*}{*|},
moredelim=*[is][\itshape\rmfamily]{/*}{*/},
moredelim=[is][\ttfamily\color{dkviolet}]{\{-}{-\}}
}

\definecolor{dkblue}{rgb}{0,0.1,0.5}
\definecolor{lightblue}{rgb}{0,0.5,0.5}
\definecolor{dkgreen}{rgb}{0,0.4,0}
\definecolor{dk2green}{rgb}{0.4,0,0}
\definecolor{dkviolet}{rgb}{0.6,0,0.8}
\definecolor{dkpink}{rgb}{0.75,0,1}\definecolor{firebrick}{rgb}{0.69,0.13,0.13}

\newif\ifelec

\electrue
\definecolor{burntorange}{rgb}{0.8, 0.33, 0.0}

\lstdefinelanguage{SQLL}{
mathescape=true,
texcl=false,
morekeywords=[1]{select, from, where, group, by, having, create, table, insert, into, values, distinct, order},
morekeywords=[2]{exists, any, all, true, false, unknown, not, in},
morekeywords=[3]{sum, SUM, avg, AVG, count, COUNT, min, MIN, max, MAX, integer, NULL},
morekeywords=[4]{
         exists2,
         pose, set, move, case, elim, apply, clear,
            hnf, intro, intros, generalize, rename, pattern, after,
	    destruct, induction, using, refine, inversion, injection,
            constructor,
         rewrite, congr, unlock, compute, vm_compute, native_compute,
            replace, fold, unfold, change, cutrewrite, simpl,
            cbv, lazy,
         have, suff, wlog, suffices, without, loss, nat_norm,
            assert, cut, trivial, revert, bool_congr, nat_congr,
	 symmetry, transitivity, auto, split,
         autorewrite,
       interval_intro},
morekeywords=[5]{
         done, exact, reflexivity, tauto, romega, omega,
         assumption, solve, contradiction, discriminate,
         ring, field, interval},
morekeywords=[6]{do, last, first, try, idtac, repeat, progress},
literate=
        {:=}{{$\mathrel{\mathop:\mathopen=}$}}2
{=}{{$=$}}1
{==}{{$\equiv$}}1
        {!=}{{$\not\equiv$}}1
        {<=}{{$\leq$}}1
        {>=}{{$\geq$}}1
        {<>}{{$\neq$}}1
        {->}{{$\rightarrow$}}2
{<-}{{$\leftarrow$}}2
{=>}{{$\Rightarrow$}}1
{/\\}{{$\wedge$}}2
        {\\/}{{$\vee$}}2
        {<->}{{$\leftrightarrow$}}2
        {<=>}{{$\Leftrightarrow$}}2
{forall\ }{{$\forall\,$}}1
{negb}{{$\neg$}}1
        {~}{{$\neg$}}1
        {\\in}{{$\in$}}1
{\$}{{\textcolor{blue}{\$}}}1
        {^-1}{{$^{-1}$}}1,
comment=[f][\color{red}][0]{--},
showstringspaces=false,
morestring=[b]",
morestring=[d]´,
extendedchars=true,
sensitive=true,
breaklines=true,
basicstyle=\footnotesize\ttfamily,
captionpos=b,
columns=[l]fullflexible,
keepspaces=true,
identifierstyle={\ttfamily\ifelec\color{black}\fi},
keywordstyle=[1]{\ttfamily\ifelec\color{dkviolet}\fi},
keywordstyle=[2]{\ttfamily\ifelec\color{dkgreen}\fi},
keywordstyle=[3]{\ttfamily\ifelec\color{burntorange}\fi},
keywordstyle=[4]{\ttfamily\ifelec\color{dkblue}\fi},
keywordstyle=[5]{\ttfamily\ifelec\color{red}\fi},
keywordstyle=[6]{\ttfamily\ifelec\color{dkpink}\fi},
stringstyle=\ttfamily,
commentstyle={\ttfamily\ifelec\color{firebrick}\fi},
moredelim=[is][\color{burntorange}\bfseries\ttfamily]{|*}{*|},
moredelim=*[is][\itshape\rmfamily]{/*}{*/},
moredelim=[is][\ttfamily\color{dkviolet}]{\{-}{-\}}
}

\definecolor{dkblue}{rgb}{0,0.1,0.5}
\definecolor{lightblue}{rgb}{0,0.5,0.5}
\definecolor{dkgreen}{rgb}{0,0.4,0}
\definecolor{dk2green}{rgb}{0.4,0,0}
\definecolor{dkviolet}{rgb}{0.6,0,0.8}
\definecolor{dkpink}{rgb}{0.75,0,1}\definecolor{firebrick}{rgb}{0.69,0.13,0.13}

\definecolor{burntorange}{rgb}{0.8, 0.33, 0.0}
\definecolor{antiquefuchsia}{rgb}{0.57, 0.36, 0.51}
\definecolor{pomegranate}{RGB}{192, 57, 43}
\definecolor{light-gray}{gray}{0.95}
\definecolor{ivory}{RGB}{239,239,234}\definecolor{mydarkred}{rgb}{0.8,0,0.}

\newcommand{\Sql}{\lstinline[language=SQLL,basicstyle=\tt\small,mathescape=true]}
\newcommand{\Sqll}{\lstinline[language=SQLL,mathescape=true]}

\lstnewenvironment{coqlisting}{\lstset{language=Coq,fontadjust=true,columns=fullflexible,showspaces=false}
\lstset{basicstyle=\sf\fontsize{9}{10.69}\selectfont,commentstyle=\it\sffamily\fontsize{9}{10}\color{mydarkred},keywordstyle=\sffamily\color{blue}}
\lstset{escapeinside={(*@}{@*)}} 
}{\vspace*{-1mm plus 1mm minus 1mm}
\lstset{language=Coq,fontadjust=true,flexiblecolumns=true,showspaces=false}
\lstset{basicstyle=\sf\small,commentstyle=\it\sffamily\small\color{mydarkred},keywordstyle=\bf\sffamily\color{blue}}
\lstset{escapeinside={(*@}{@*)}} 
\lstset{commentstyle=\color{mydarkred}\it\small}
}

\lstnewenvironment{jslisting}{\lstset{language=JavaScript,fontadjust=true,columns=fullflexible,showspaces=false}
\lstset{basicstyle=\footnotesize\ttfamily,
morekeywords={fun},
commentstyle=\it\sffamily
\color{mydarkred},keywordstyle=\ttfamily\color{blue}}
\lstset{escapeinside={/*@}{@*/}}
}{\vspace*{-1mm plus 1mm minus 1mm}
\lstset{language=JavaScript,fontadjust=true,flexiblecolumns=true,showspaces=false}
\lstset{basicstyle=\footnotesize\ttfamily,
        commentstyle=\it\sffamily\small\color{mydarkred},
        keywordstyle=\bf\ttfamily\color{blue}}
\lstset{escapeinside={/*@}{@*/}}
\lstset{commentstyle=\color{mydarkred}\it\small}
}
\newcommand{\lstjs}{\lstinline[language=JavaScript,basicstyle=\tt\small]}

 \newcommand{\kkeyword}[1]{\textup{\texttt{{#1}}}}

\newcommand{\rapp}{\:@\:}

\newcommand{\qapp}{\:@\:}

\newcommand{\qor}{??}

\newcommand{\qmap}[2]{\chi_{\left\langle{\scriptstyle#1}\right\rangle}\!\left({#2}\right)}
\newcommand{\qmapenv}[1]{\chi^e_{\left\langle{\scriptstyle#1}\right\rangle}}

\newcommand{\qselect}[2]{\sigma\!_{\left\langle{\scriptstyle#1}\right\rangle}\!\left({#2}\right)}
\newcommand{\qeither}[2]{{#1} | {#2}}
\newcommand{\qID}{\kkeyword{In}}
\newcommand{\qENV}{\kkeyword{Env}}

\newcommand{\qgroupby}[3]{\ensuremath{\kkeyword{group\_by}_{#1}({#2}, {#3})}}

\newcommand{\opunop}{\boxplus}
\newcommand{\opbinop}{\boxtimes}

\newcommand{\opconcat}{\oplus}

\newcommand{\true}{\ensuremath{\kkeyword{true}}\xspace}
\newcommand{\false}{\ensuremath{\kkeyword{false}}\xspace}
\newcommand{\dleftOp}{\ensuremath{\kkeyword{left}}\xspace}
\newcommand{\drightOp}{\ensuremath{\kkeyword{right}}\xspace}
\newcommand{\dleft}[1]{\ensuremath{\kkeyword{left}~{#1}}\xspace}
\newcommand{\dright}[1]{\ensuremath{\kkeyword{right}~{#1}}\xspace}
\newcommand{\dunit}{\ensuremath{\kkeyword{()}}\xspace}

\newenvironment{gram}{
  \[
  \begin{array}{lcl}
}{
  \end{array}
  \]
}

\newcommand{\Downarrowa}{\Downarrow_a}

\newcommand{\envnraenv}{\ensuremath{\rho}}
\newcommand{\envimp}{\ensuremath{\rho}}

\newcommand{\typeeither}{\ensuremath{\mathtt{either}}}

\newcommand{\impassign}[2]{\ensuremath{{#1}~\texttt{:=}~{#2}}}
\newcommand{\impforrange}[3]{\ensuremath{\texttt{for}~{#1}\
    \texttt{in}~{#2}\ \texttt{do}\ {#3}}}
\newcommand{\impif}[3]{\ensuremath{\texttt{if}~{#1}\ \texttt{then}\
    {#2}\ \texttt{else}\ {#3}}}
\newcommand{\impfun}[3]{\ensuremath{\texttt{fun(}{#1}\texttt{)\{}\
    {#2}\texttt{;}\ \texttt{return}\ {#3}\texttt{\}\;}}}
\newcommand{\impblock}[1]{\ensuremath{\texttt{\{}\ {#1}\ \texttt{\}}}}
\newcommand{\impvar}[1]{\ensuremath{\texttt{var}\ {#1}}}
\newcommand{\impvarinit}[2]{\ensuremath{\texttt{var}\ {#1}\ \texttt{=}\ {#2}}}

\newcommand{\algoif}[3]{\ensuremath{\mathit{if}~{#1}\ \mathit{then}\
    {#2}\ \mathit{else}\ {#3}}}
\newcommand{\algofold}[3]{\ensuremath{\mathit{fold}~{#1}\ {#2}\ {#3}}}

\newcommand{\tobool}[1]{\ensuremath{\mathit{toBool}({#1})}}
\newcommand{\tolist}[1]{\ensuremath{\mathit{toList}({#1})}}

\newcommand{\impsem}[2]{\ensuremath{\llbracket#1\rrbracket^{\mbox{\tiny imp}}(#2)}}
\newcommand{\impdatasem}[1]{\ensuremath{\{\mkern-3.8mu[ #1 ]\mkern-3.8mu\}}}

 \newcommand{\dbcert}{\mbox{DBCert}\xspace}
\newcommand{\alasql}{\mbox{AlaSQL}\xspace}
\newcommand*{\QED}{\hfill\ensuremath{\square}}

\newcommand{\Tau}{\mathcal{T}}
\newcommand{\aenv}{\ensuremath{\mathcal{A}}}

\newcommand{\tratt}[1]{\ensuremath{\Tau^{\mathsf{att}}(#1)}}
\newcommand{\trtab}[1]{\ensuremath{\Tau^{\mathsf{tab}}(#1)}}
\newcommand{\trval}[1]{\ensuremath{\Tau^{\mathsf{val}}(#1)}}
\newcommand{\trB}[1]{\ensuremath{\Tau^{\mathsf{B}}(#1)}}

\newcommand{\trbag}[1]{\ensuremath{\Tau^{\mathsf{bag}}(#1)}}
\newcommand{\tenv}[1]{\ensuremath{\Tau^{\mathsf{env\_dynamic}}(#1)}}
\newcommand{\taenv}[1]{\ensuremath{\Tau^{\mathsf{env\_static}}(#1)}}
\newcommand{\trft}[2]{\ensuremath{\Tau^{\mathsf{\ef}}_{\scriptstyle#1}(#2)}}
\newcommand{\tra}[2]{\ensuremath{\Tau^{\mathsf{\ea}}_{\scriptstyle#1}(#2)}}
\newcommand{\tre}[2]{\ensuremath{\Tau^{\mathsf{e}}_{\scriptstyle#1}(#2)}}
\newcommand{\trfn}[2]{\ensuremath{\Tau^{\mathsf{\fn}}(#1,#2)}}
\newcommand{\trp}[2]{\ensuremath{\Tau^{\mathsf{\pr}}(#1,#2)}}
\newcommand{\trag}[2]{\ensuremath{\Tau^{\mathsf{\ag}}(#1,#2)}}
\newcommand{\trq}[2]{\ensuremath{\Tau^{\mathsf{\algquery}}_{#1}(#2)}}
\newcommand{\trf}[2]{\ensuremath{\Tau^{\mathsf{f}}_{\scriptstyle#1}(#2)}}
\newcommand{\trsel}[2]{\ensuremath{\Tau^{\mathsf{Sel}}_{\scriptstyle#1}(#2)}}
\newcommand{\tri}[1]{\ensuremath{\Tau^{\mathsf{i}}({#1})}}

\newcommand{\eapp}[2]{\ensuremath{#2~\circ^e~#1}}
\newcommand{\data}{\texttt{data}\xspace}
\newcommand{\tdot}[2]{\ensuremath{#1\boldsymbol{\cdot}#2}}

\newcommand{\pushd}{\textsf{\mbox{push$_{\mathit{one}}$}}}
\newcommand{\pushq}{\textsf{\mbox{push$_{\mathit{bag}}$}}}
\newcommand{\dbag}[1]{[ {#1} ]\xspace}

\newcommand{\findevalaenv}[2]{\ensuremath{{\mathbb F}_{a}(#1,#2)}}

\begin{document}

\title[Translating Canonical SQL to Imperative Code in Coq]{Translating Canonical SQL to Imperative Code in Coq}

\author{Véronique Benzaken}
\affiliation{
  \position{Université de Paris Saclay}
\institution{LMF, Université Paris-Saclay}            \country{France}                    }
\email{veronique.benzaken@universite-paris-saclay.fr}          \author{Évelyne Contejean}
\affiliation{
  \position{CNRS - Université de Paris Saclay}
\institution{LMF, CNRS, Université Paris-Saclay}            \country{France}                    }
\email{Evelyne.Contejean@lri.fr}          

\author{Mohammed Houssem Hachmaoui}
\affiliation{
  \position{Université de Paris Saclay}
\institution{LMF, Université Paris-Saclay}            \country{France}                    }
\email{mohammed.hachmaoui@lri.fr}          

\author{Chantal Keller}
\affiliation{
  \position{Université de Paris Saclay}
\institution{LMF, Université Paris-Saclay}            \country{France}                    }
\email{Chantal.Keller@lri.fr}          

\author{Louis Mandel}
\affiliation{
  \position{IBM Research}
\institution{IBM Research}            \country{USA}                    }
\email{lmandel@us.ibm.com}          

\author{Avraham Shinnar}
\affiliation{
  \position{IBM Research}
\institution{IBM Research}            \country{USA}                    }
\email{shinnar@us.ibm.com}

\author{Jérôme Siméon}
\authornote{This author's work conducted while at Clause, Inc.}          \affiliation{
  \position{Scientist}
\institution{DocuSign, Inc.}            \country{USA}                    }
\email{jerome.simeon@docusign.com}
 
\begin{abstract}
  SQL is by far the most widely used and implemented query language. Yet, on some key features, such as correlated queries and \null value semantics, many implementations diverge or contain bugs.
  We leverage recent advances in the formalization of SQL and query compilers to develop \dbcert, the first mechanically verified compiler from \sql queries written in a canonical form to imperative code.
  Building \dbcert required several new contributions which are described in this paper.
  First, we specify and mechanize a complete translation from SQL to the Nested Relational Algebra which can be used for query optimization.
  Second, we define Imp, a small imperative language sufficient to express SQL and which can target several execution languages including JavaScript.
  Finally, we develop a mechanized translation from the nested relational algebra to Imp, using the nested relational calculus as an intermediate step.
\end{abstract}



\sloppy

\maketitle

\renewcommand{\shortauthors}{V. Benzaken, É. Contejean, M. H. Hachmaoui,
  C. Keller, L. Mandel, A. Shinnar, and J. Siméon}

\section{Introduction}\label{sec:intro}

SQL is by far the most widely used query language. While originally
designed to query relational databases, it is now also used for data
integration~\cite{DBLP:conf/icitcs/LeeCKSW16}, for processing
logs~\cite{jin2010streamsql}, and for big
data~\cite{DBLP:conf/bigdataconf/GroverGJYCMM15}. It is commonly
available as a library in a number of programming
languages~\cite{alasql,sqlalchemy}.

While the SQL semantics for flat select-project-join queries are well
understood and consistent across platforms, several important features
such as nested queries (often called correlated queries in the
database literature) and \null values are a common source of
bugs~\cite{DBLP:journals/pvldb/GuagliardoL17,DBLP:conf/cpp/BenzakenC19}.
Since SQL is commonly used in critical applications and for handling
sensitive data, like accessing medical information, SQL
implementations can benefit from the use of formal verification
techniques.

Despite recent progress in mechanized semantics for query
languages in general~\cite{DBLP:conf/ecoop/ShinnarSH15} and for SQL in
particular~\cite{Chu:2017:HPQ:3062341.3062348,DBLP:conf/cpp/BenzakenC19},
those are far from being usable as a \sql implementation, typically
missing a query optimizer and the ability to generate efficient code.
In this paper, we describe \dbcert, a compiler from \sql to imperative
code which addresses those limitations and is mechanically verified
using the Coq proof assistant.

To build \dbcert, we followed a classical database compiler
architecture~\cite{DBLP:conf/sigmod/ShaikhhaKPBD016}: (1)~a source language to write the queries, (2)~an
algebra suitable for optimization, and (3)~a physical plan to execute the queries, specific to the targeted runtime.
Our compiler enhances this architecture by formally defining each of these components
and proving the translation between them correct.

For the source language~(1), we use the mechanized \sql semantics
from~\citet{DBLP:conf/cpp/BenzakenC19}.  To our knowledge, this is the
most complete formal semantics of \sql currently available. That
semantics is fully executable and supports a large subset of \sql,
including: \Sql{select from where group by having} blocks, \null
values, aggregate functions, and correlated queries.\footnote{The
  constructs that are not handled yet which represent a loss of
  expressiveness are: silent coercion from singleton bags to values,
  recursive queries, \Sql{distinct}, and \Sql{order by}.}

For the intermediate algebra~(2), we use the Nested Relational
Algebra~(\nrae) from~\citet{DBLP:conf/sigmod/AuerbachHMSS17} which
comes with a Coq mechanization, including a multi-step, optimizing
compiler. \nrae has several features essential to capture SQL
queries. First it can naturally handle nested queries. Second, it
includes operators over sum types which we use to encode the semantics
of \null values. Finally, \nrae was chosen for its validated use for optimization of nested
queries~\cite{DBLP:conf/dbpl/CluetM93,claussen1997optimizing,querycompilers}
and \null values~\cite{DBLP:journals/tkde/ClaussenKMPS00}.

For the physical plan~(3), we generate \js code
to target an in-memory database where the data are represented as JSON
objects. We chose \js for our final output because it is very
portable.
For the formalization, we defined a series of intermediate languages
that progressively change to programming model from \nrae to a small
imperative language \imp, which is sufficiently expressive to capture
SQL semantics. \imp is parameterized by a data model and a set of
operators which makes it flexible enough to drive code generation for
a range of target execution languages.

\paragraph*{Architecture}

The following diagram outlines the full \dbcert compilation pipeline.

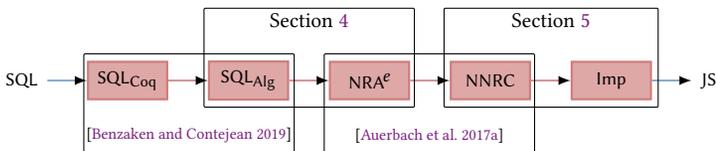
\begin{center}
  \begin{minipage}[c]{.9\linewidth}
    \centering
    \newcommand{\coqHTMLBase}{https://querycert.github.io}
\newcommand{\coqBaseModule}{html}
\newcommand{\coqurl}[3]{\coqHTMLBase/\coqBaseModule/Qcert.#1.#2.html\##3}

\tikzset{
    hyperlink node/.style={
        alias=sourcenode,
        append after command={
            let \p1 = (sourcenode.north west),
                \p2=(sourcenode.south east),
                \n1={\x2-\x1},
                \n2={\y1-\y2} in
            node [inner sep=0pt, outer sep=0pt,anchor=north west,at=(\p1)] {\href{#1}{\XeTeXLinkBox{\phantom{\rule{\n1}{\n2}}}}}
}
    }
}

\definecolor{cjava}{RGB}{156,186,95}
\definecolor{ccoqp}{RGB}{222,168,167}
\definecolor{ccoqc}{RGB}{161,187,215}

\tikzstyle{coqp}=[
 draw=ccoqp!150,fill=ccoqp
]

\tikzstyle{coqc}=[
 draw=ccoqc!150,fill=ccoqc
]

\tikzstyle{java}=[
 draw=cjava!150,fill=cjava
]

\tikzstyle{java-api}=[
 draw=cjava,fill=cjava!50,
 minimum width=21em,
 minimum height=1.4em,
 xshift=5em,
 font=\scriptsize\sffamily,
]

\tikzstyle{source}=[
 text width=2.75em,
 align=right
]

\tikzstyle{target}=[
 text width=4em,
 align=left
]

\tikzstyle{lang}=[
 thick,
 minimum width=3em,
 minimum height=1.4em,
 font=\scriptsize\sffamily,
]

\tikzstyle{sep-caption}=[
 thick,
 font=\scriptsize\sffamily,
]

\tikzstyle{tcoqp}=[
 draw=ccoqp!150
]

\tikzstyle{tcoqc}=[
 draw=ccoqc!150
]

\tikzstyle{trans}=[
 draw,semithick,-latex,
 font=\scriptsize\sffamily,
]

\tikzstyle{separation}=[
 draw,very thick, gray
]

\newcommand{\tikzoptim}[3][]{
   \path[trans, draw=#1!150, fill=#1!150] (#2) edge [looseness=10, loop above]
       node [pos=0.5](#2-optim) {}
       node [yshift = 0.5em,below] {~} ();
}
\newcommand{\tikzoptimdotted}[3][]{
   \path[trans, draw=#1!150, fill=#1!150] (#2) edge [looseness=10, loop above, dotted]
       node [pos=0.5](#2-optim) {}
       node [yshift = 0.5em,below] {{#3}} ();
}

\newcommand{\transref}[2]{
}

\newcommand{\transocamljava}[1]{
 \path[draw,thick,latex-latex, densely dotted,
      shorten >=0.2em, shorten <=0.2em,
       java] ([xshift=-0.75em] #1.south) -- ([xshift=-0.75em] #1-java.north);
}

 \begin{tikzpicture}[
   align=center,
   node distance=1.25em and 1.5em,
   font=\Huge,
   every loop/.style={latex-},
 ]

 \node[lang, source,
       ]
      (sql) {\sql};

 \node[lang, coqp,
       right=of sql]
      (sqlcoq) {\sqlcoq};

 \node[lang, coqp,
       right=of sqlcoq]
      (sqlalg) {\sqlalg};

 \node[lang, coqp,
       right=of sqlalg]
      (nraenv) {\NRAEnv};

 \node[lang, coqp,
       right=of nraenv]
      (nnrc) {NNRC};

 \node[lang, coqp,
       right=of nnrc]
      (imp) {Imp};

 \node[lang, target,
       right=of imp]
      (js) {JS};

 \path[trans, tcoqc] (sql.east) -- (sqlcoq.west)
   \transref{3}{\coqurl{Translation}{NRAEnvtoNNRC}{nraenv_to_nnrc_top}};

 \path[trans, tcoqp] (sqlcoq.east) -- (sqlalg.west)
   \transref{3}{\coqurl{Translation}{NRAEnvtoNNRC}{nraenv_to_nnrc_top}};

 \path[trans, tcoqp] (sqlalg.east) -- (nraenv.west)
   \transref{3}{\coqurl{Translation}{NRAEnvtoNNRC}{nraenv_to_nnrc_top}};

 \path[trans, tcoqp] (nraenv.east) -- (nnrc.west)
   \transref{3}{\coqurl{Translation}{NRAEnvtoNNRC}{nraenv_to_nnrc_top}};

 \path[trans, tcoqp] (nnrc.east) -- (imp.west)
   \transref{3}{\coqurl{Translation}{NRAEnvtoNNRC}{nraenv_to_nnrc_top}};

 \path[trans, tcoqc] (imp.east) -- (js.west)
   \transref{3}{\coqurl{Translation}{NRAEnvtoNNRC}{nraenv_to_nnrc_top}};

   \node[draw,minimum width=2.8cm,minimum height=1.3cm,
   rounded corners=.5pt,label={[label distance=-1.3cm]90:{{\tiny\cite{DBLP:conf/cpp/BenzakenC19}}}}] (cpp)
   at (2.5,-.3) {};  
   \node[draw,minimum width=2.8cm,minimum height=1.3cm,
   rounded corners=.5pt,label={[label distance=-.38cm]90:{{\footnotesize \Cref{sec:sqlatonrae}}}}] (sqlatonrae)
   at (4.1,.3) {};   
   \node[draw,minimum width=2.8cm,minimum height=1.3cm,
   rounded corners=.5pt,label={[label distance=-1.3cm]90:{{\tiny\cite{DBLP:conf/sigmod/AuerbachHMSS17}}}}] (sigmod)
   at (5.7,-.3) {};     
   \node[draw,minimum width=2.85cm,minimum height=1.3cm,
   rounded corners=.5pt,label={[label distance=-.38cm]90:{{\footnotesize \Cref{sec:nnrctoimp}}}}] (nnrctoimp)
   at (7.32,.3) {};

\end{tikzpicture}

   \end{minipage}
\end{center}

The compilation from \sqlcoq, a canonical form for \sql queries, to \imp is fully verified: the semantics
of any valid \sqlcoq query is preserved by the compilation.
\sqlcoq is first compiled to \sqlalg,
an extension of the relational algebra that includes a \sql grouping
operator, formulas and environment handling.
That algebra is translated to \nrae, which is used for query
optimization.
Finally, \nrae is translated to the small imperative language
\imp. This translation is decomposed into multiple steps using different
intermediate languages, starting with the Named Nested Relational
Calculus (\nnrc)~\cite{BusscheV07}, a functional language with list
comprehensions.
Using \nnrc is a pragmatic choice, as it is proven
equivalent to \nrae~\cite{DBLP:conf/sigmod/AuerbachHMSS17} and is closer
to traditional languages, having
variable names instead of just using combinators.

To make the resulting compiler usable, it is complemented by
non-verified front- and back-ends: a parser for \sql into \sqlcoq,
which also performs simple disambiguation ({\em e.g.} to avoid name
clashes) and a code generation step from \imp to \js.  The generated
code by the compiler linked to a runtime can then be used as a Node.js
library.

\paragraph*{Contributions}

This paper makes the following contributions:
\begin{itemize}
\item A translation from \sql to the optimizing algebra \nrae. To the
  best of our knowledge, there is no description~(including in
  database literature) of such a translation for such a large fragment
  of \sql, and we will see that this step is not trivial.
This part represents about 20,000 lines of new formalization and
  proofs.
\item A translation from \nrae to a physical plan for an in-memory database.
  This includes the definition of multiple intermediate languages,
  introduced to break the difficulty of the proofs, and the language
  \imp, a simple imperative language sufficiently expressive to capture
  SQL semantics. In addition to the resulting correct-by-construction
  back-end, this is an advance in verification techniques.
This part represents around 31,000 lines of new formalization and proofs.
\item A complete compiler for a large subset of SQL, written in Coq,
  that translates to a database algebra suitable for optimization, and
  generates low-level imperative code for execution. This bridges a
  gap between prior works on \sql formalization and on mechanization
  of query compilers~\citep{popl10morisset,
    DBLP:conf/sigmod/AuerbachHMSS17a}.
\end{itemize}

While we rely heavily on prior work, building \dbcert required
significant new development. First, we had to bridge the gap
between the source \sql semantics and the algebraic intermediate
representation in \nrae.
In particular, \nrae does not have a builtin for $\null$ and must have an explicit encoding of the \sql environments semantics.
Second, we had to develop translations and correctness proofs from the algebraic representation to a lower-level imperative language.
This proof necessitated the creation of a series of intermediate languages to cope with its complexity.

\paragraph*{Outline}

\Cref{sec:overview} presents some simple \sql examples illustrating
subtleties of the semantics of the language (\Cref{subsec:examples}) and then explains the compilation pipeline on an example (\Cref{subsec:sql-to-js}).
\Cref{sec:preliminaries} defines the main languages used in the
compiler.
\Cref{sec:sqlatonrae} describes the translation from \sqlalg to \nrae,
This translation handles delicate aspects of \sql, including \null
values and environments for correlated queries.
\Cref{sec:nnrctoimp} describes the translation from \nnrc, an
expression oriented functional language, to \imp, a statement oriented
imperative language with mutable variables.
\Cref{sec:implementation} reviews the \dbcert implementation. The
compiler is verified using the Coq proof assistant and extracted to
OCaml. The non-verified parts include the \sql parser, written
in OCaml, and the \js code generation and runtime used for execution.
\Cref{sec:related} evaluates the compiler on some challenging queries
and discusses methodology.
This paper provides insight both on the compilation of \sql, and on
the software engineering aspect of connecting two large Coq projects
developed independently.

\ifextended
This article is an extended version with appendix of the one published at OOPSLA~2022~\cite{oopsla22}.
\fi{}
This article is also accompanied with an artifact~\cite{artifact} which is a Docker image containing an installed version of a snapshot of the following open source projects:
\begin{itemize}
\item \url{https://github.com/dbcert/dbcert}: the entry point of the project containing the main theorem and providing the runtime and the runner
\item \url{https://framagit.org/formaldata/sqltonracert}: the
  frontend from SQL to \nrae (which itself relies on the SQL formal semantics)
\item \url{https://querycert.github.io}: the Q*Cert compiler that contains in particular the backend from \nrae to \imp.
\end{itemize}

\section{Overview}\label{sec:overview}

\subsection{Challenges of the SQL semantics}\label{subsec:examples}
As an introductory example, let us consider the following SQL query
written using the \alasql~\cite{alasql} library for Node.js:

\begin{lstlisting}[language=SQLL]
alasql('CREATE TABLE R (a number, b number)');
alasql.tables.R.data = [ {a: 1, b: 10}, {a: 2, b: 20}, {a: 3, b: 30} ];
var res = alasql('select a from R where b > 15'); // res = [ { "a": 2 }, { "a": 3 } ]
\end{lstlisting}

This library is primarily used for querying JSON data in memory or to
run \sql queries directly in the browser. The first line declares a
relational schema with one table \Sql!R!  containing two columns
\Sql!a! and \Sql!b!, both of type number. The second line populates the
database, here as a JavaScript array of objects, where each object
corresponds to a row in table \Sql!R! with those same fields \Sql!a!
and \Sql!b!. The third line executes a simple \Sql{select from where}
statement which returns the \Sql!a! column for every row which has a
\Sql!b! column greater than 15.

\paragraph{Challenges with null value semantics}

We next consider a query adapted
from~\citet{DBLP:journals/pvldb/GuagliardoL17} involving \null values, an important feature of \sql commonly used to model missing data.
This query selects all the values in the table \Sql{R} that are for
sure not in \Sql{S}.
\begin{lstlisting}[language=SQLL]
alasql('CREATE TABLE R (a number)');
alasql('CREATE TABLE S (b number)');
alasql.tables.R.data = [ {a: 1}, {a: null} ];
alasql.tables.S.data = [ {b: null} ];
var res = alasql('select a from R where a not in (select b from S)');
// expected: res = []
// alasql:   res = [ {"a": 1} ]
\end{lstlisting}

The nested query \Sql!select b from S! returns a table with a single row
containing \Sql!null!. For each row in \Sql!R!, the value of the
attribute \Sql!a! is compared to \Sql!null! to test non-membership in
the result of \Sql!select b from S!. The  \Sql!not in!
predicate expands to \Sql!1 <> null! for the first row and
\Sql!null <> null! for the second row.
\sql uses a three-valued logic and comparing a value to \null returns
\emph{unknown}.  Thus, the two comparisons return \emph{unknown}
and the query's result should be the empty collection.

Unfortunately, \alasql, when given this query, incorrectly returns \Sql![{a:1}]! instead.
\alasql probably relies on the
JavaScript comparison where \Sql!1 <> null! is $\mathit{true}$ instead
of \emph{unknown}.

\paragraph{Challenges with nested query semantics}

A main challenge when compiling \sql to \js is correctly handling nested queries.  In particular, extra care needs to be taken to account for correlated queries, where the inner query refers to a variable introduced by an outer query.

Correlated queries are an important feature of \sql since they allow in particular to answer negative questions like the previous query~(where the correlation is introduced by the \Sql{in} operator).

To illustrate the subtlety of the semantics of nested queries, we consider the following queries \Sql{Q1} and \Sql{Q2}.  The only difference between them is in the expression \Sql{sum(1+0*$b$)}: the variable $b$ refers to \Sql{b2} (defined in table \Sql{t2}) in \Sql{Q1} and to \Sql{b1} (defined in table \Sql{t1}) in \Sql{Q2}.
\medskip

\noindent
\begin{minipage}[c]{.615\linewidth}
\begin{lstlisting}[language=SQLL]
-- Q1
select a1 from t1 group by a1 having exists
  (select a2 from t2 group by a2 having sum(1+0*b2) = 2);
\end{lstlisting}
\begin{lstlisting}[language=SQLL]
-- Q2
select a1 from t1 group by a1 having exists
  (select a2 from t2 group by a2 having sum(1+0*b1) = 2);
\end{lstlisting}
\end{minipage}
$\,\,\,$
\begin{minipage}[c]{.36\linewidth}
\footnotesize
\def\arraystretch{0.8}
\begin{tabular}{cc c cc c c c c}
\toprule
\multicolumn{2}{c}{\Sqll{t1}} && \multicolumn{2}{c}{\Sqll{t2}} && \multicolumn{1}{c}{\Sqll{Q1}} && \multicolumn{1}{c}{\Sqll{Q2}} \\
\cmidrule(lr){1-2}\cmidrule(lr){4-5}\cmidrule(lr){7-7}\cmidrule(lr){9-9}
\Sqll!a1! & \Sqll!b1! && \Sqll!a2! & \Sqll!b2! && \Sqll!a1! && \Sqll!a1! \\\midrule
      1   &       1   &&       7   &       7   &&       1   &&       1   \\
      1   &       2   &&       7   &       8   &&           &&           \\
      2   &       3   &&           &           &&       2   &&           \\
      3   &       1   &&           &           &&       3   &&           \\
      3   &       2   &&           &           &&           &&           \\
      3   &       3   &&           &           &&           &&           \\
\bottomrule
\end{tabular}
\end{minipage}
\medskip

At first glance, adding a term equal to $0$ in a
sum should have no effect.  However, \Sql{sum} is an aggregate operator, and
is executed on each element of the table containing~$b$.  Thus, the expression
\Sql{sum(1+0*$b$)} effectively counts the number of occurrences of $b$.
This expression therefore returns different results when applied to table \Sql{t1} (\Sql{sum(1+0*$b1$)}, as in Q2) and \Sql{t2} (\Sql{sum(1+0*$b2$)}, as in Q1).

The semantics of \Sql{t1 group by a1} is to split the table \Sql{t1} into intermediate tables where the values of the attribute \Sql{a1} is the same. In our example, for both queries, it creates the tables \Sql![{a1:1, b1:1}, {a1:1, b1:2}]!, \Sql![{a1:2, b1:3}]!, and \Sql![{a1:3, b1:1}, {a1:3, b1:2}, {a1:3, b1:3}]!.
Then, on each of these tables, the condition \Sql{having} is executed.
The expression \Sql{select a2 from t2 group by a2 having sum(1+0*$b$) = 2} is thus executed three times in three different contexts.
The expression \Sql{t2 group by a2} always creates the table \Sql![{a2:7, b2:7}, {a2:7, b2:8}]! and
the condition \Sql{having sum(1+0*$b$) = 2} tests if the number of occurrences of~$b$ is two.

For \Sql{Q1}, where $b =$~\Sql{b2}, since there are three times the condition is true (since the table containing \Sql{b2} has two elements), the inner query returns \Sql![{a2:7}]!, and the \Sql{exists} condition is always a success.  As a result, the outer query returns \Sql![{a1:1}, {a1:2}, {a1:3}]!.

For \Sql{Q2}, where $b =$~\Sql{b1}, the table containing \Sql{b1} has two elements such that \Sql{a1 = 1}, one element such that \Sql{a1 = 2}, and three elements such that \Sql{a1 = 3}.
The condition is true only once (when \Sql{a1 = 1}), so the inner query returns \Sql![{a2:7}]! and the \Sql{exists} condition succeeds only in this case.
Thus, the outer query returns \Sql![{a1:1}]!.
\alasql, alas, produces an incorrect result for this query.

This example illustrates what we call the {\em environment handling} of
SQL: evaluating a nested query must be done in an environment aware of
all outer queries, and one must be careful on correctly choosing the
important piece of information in this environment. We will detail this
in Sections~\ref{sec:preliminaries:sqlalg} and
\ref{sec:sqlatonrae:translation:envtran}.

\medskip

\subsection{Translating SQL to JavaScript}\label{subsec:sql-to-js}

We introduce our translation using a simple correlated query that returns all the
values of the attribute \Sql{a} of the table \Sql{R} which are present in the column \Sql{b} of \Sql{S}:
\begin{lstlisting}[language=SQLL]
select a from R where exists (select b from S where b = a);
\end{lstlisting}

We first translate the query into \sqlcoq, a subset of \sql where all implicit features of \sql are explicit.
For example, a \Sql{select} without a \Sql{where} clause is completed by \Sql{where true}.
\sqlcoq is as expressive as the considered subset of \sql, but its regularity simplifies formalization.
Our example in \sqlcoq becomes:
\begin{lstlisting}[language=SQLL]
select x as a from (table R) t0(x)
  where exists (select y as t1_y from (table S) t1(y) where y = x);
\end{lstlisting}
In \sqlcoq, all the intermediate results must be named.
For example, the notation \Sql{(table R) t0(x)} renames the table~\Sql{R} into~\Sql{t0} and its single attribute is renamed~\Sql{x}.
The selection \Sql{x as a} projects the attribute \Sql{x} of the table \Sql{t0} and renames it \Sql{a}. The definition of \sqlcoq and this compilation step is taken from~\citet{DBLP:conf/cpp/BenzakenC19}.
Currently, the \sql features not supported by \sqlcoq are silent
coercion from singleton bags to values, recursive queries,
\Sql{distinct}, and \Sql{order by}.

\medskip

The next compilation step is also taken from~\citet{DBLP:conf/cpp/BenzakenC19}.
It translates \sqlcoq to \sqlalg, a relational algebra such as is found in database
textbooks~\cite{DBLP:books/cs/Ullman82,DBLP:books/aw/AbiteboulHV95}, 
but including grouping and aggregates.
\sqlalg includes operators such as projection~${\large \pi}$, selection~${\large \sigma}$, natural join~${\large \bowtie}$, and a grouping operator~${\large \gamma}$.
Our example query translated into \sqlalg is:
$$
{\large\pi}_{x \mathtt{~as~} a}
 ({\large\sigma}_{
  \mathtt{exists}
   ({\large\pi}_{\scriptstyle y \mathtt{~as~} t1\_y}
    ({\large\sigma}_{\scriptstyle y = x}
     ({\large \pi}_{\scriptstyle b \mathtt{~as~} y}(S))))}
  ({\large\pi}_{a \mathtt{~as~} x}(R))
$$
The input of this query is the expression ${{\large \pi}_{a \mathtt{~as~} x}(R)}$, corresponding to \Sql{(table R) t0(x)}, the renaming of the attribute of the table~\Sql{R}.
The top-level ${\large\pi}_{x \mathtt{~as~} a}$ corresponds to the projection of the result by the clause \Sql{select x as a}.
It is applied to ${\large\sigma}_{\mathtt{exists}(...)}$, which corresponds to the \Sql{where exists (...)} clause.
Similarly, the expression inside the $\mathtt{exists}$ predicate corresponds to the inner \sqlcoq query.

\medskip

From \sqlalg, the query is translated into \nrae~\cite{DBLP:conf/sigmod/AuerbachHMSS17}, a nested relational algebra.
This intermediate language has two purposes: (1)~it makes explicit the encoding of \sql features like \null values and the environment handling, and (2)~it is a good language for optimization~\cite{DBLP:journals/tods/CaoB07,querycompilers}.
\nrae is based on functional combinators, evaluated in a context with exactly two variables: $\qID$ for the current input and $\qENV$ for the local environment.
The intuition for that translation is that the structure of relational algebra operators (e.g., ${\large \pi}$, ${\large \sigma}$) is preserved, but ``administrative'' steps
are added to deal with \null values and the \sql evaluation context is encoded in the \nrae environment $\qENV$.

Consider first the translation of ${\large\pi}_{a \mathtt{~as~} x}(R)$ from \sqlalg to \nrae:
$$
\qmap{\{ x : \qID.a \}}{R}
$$
The combinator $\chi$ is a functional map: it applies the expression
within the $\langle..\rangle$ to each element of $R$ where the element
is bound to the variable $\qID$, which holds the current input.  The expression
${\{ x : \qID.a \}}$ creates a record with label~$x$ and value the
projection of the label~$a$ from the current input~($\qID$).  As
expected, this expression creates a collection of records with
label~$x$ containing the elements of~$R$ with label~$a$.

We next consider the translation of
${\large\sigma}_{\mathtt{exists}(Q)}({\large\pi}_{a \mathtt{~as~} x}(R))$
where~$Q$ has~$S$ as input, but also depends on~$x$, the result
of~${{\large\pi}_{a \mathtt{~as~} x}(R)}$.  Denoting the translation of $Q$ as $q$, 
${\large\sigma}_{\mathtt{exists}(Q)}({\large\pi}_{a \mathtt{~as~} x}(R))$
is:
$$
\qselect{\eapp{\pushd}{\mathtt{exists}(q)}}
{\qmap{\{ x : \qID.a \}}{R}}
$$
The selection operator~${\large\sigma}$ of \sqlalg is translated into
the corresponding operator in \nrae.  But in \sqlalg,
the~${\large\sigma}$ operator implicitly adds~$x$ to the evaluation
context of~$\mathtt{exists}(q)$. This is done explicitly in \nrae,
with $\eapp{q_2}{q_1}$, which evaluates~$q_2$ first and
then evaluates~$q_1$ in the environment $\qENV$ where the result
of~$q_2$ is stored. Here, $\pushd$ adds
the value of~$x$ onto a stack defining the evaluation context
of~$\mathtt{exists}(q)$, implemented as a linked list with shape
 $\{ \mathit{slice}: \bullet, \mathit{tail}: \bullet \}$.
 $\mathit{slice}$ contains the attributes introduced by $\qmap{\{ x : \qID.a \}}{R}$ and $\mathit{tail}$ contains the evaluation context of $\qmap{\{ x : \qID.a \}}{R}$.

Finally, the last difficulties are in the translation of the
expression~${\large\sigma}_{y = x}(...)$, which has the following structure in \nrae~(to simplify the presentation we simplified the code: in particular, we assume that $y$ is not \null):
$$
\qselect{
  \eapp{\pushd}
  {(\qeither{\qID}{\mathit{false}})
    ~\circ~
   ((\qeither{\dleft{(\qENV.\mathit{slice}.y = \qID)}}{\dright{\dunit}})
   ~\circ~
   \qENV.\mathit{tail}.\mathit{slice}.x)}}{...}
$$
Starting from the right of the condition of the~${\large\sigma}$, the
$\eapp{\pushd}{}$ adds~$y$ on the top of the environment stack.  Then
$(q ~\circ~ \qENV.\mathit{tail}.\mathit{slice}.x)$ where
$q = (\qeither{\dleft{(\qENV.\mathit{slice}.y = \qID)}}{\dright{\dunit}})$
corresponds to the equality
test~$y = x$ which has to deal with \null values.
The expression $\qENV.\mathit{tail}.\mathit{slice}.x$ accesses the value of $x$ at the appropriate level in the environment stack and is given as input to $q$ using the composition operator~$\circ$.
The expression~$q$ tests if~$x$ is null~(we assume here that~$y$ is not null).
\nrae does not have a built-in notion of null, instead values that
can be \null are boxed in a value of type~$\typeeither$. For example,
the number $42$ is encoded as $\dleft{42}$ and a $\null$ is
$\dright{\dunit}$.
The operator~($\qeither{q_1}{q_2}$) matches its input with
shape $\dleft{d_1}$ and $\dright{d_2}$ to execute either $q_1$ with input~$d_1$ or $q_2$ with input~$d_2$.
In our example, if~$x$ is null, $q$~returns $\dright{\dunit}$
(corresponding to $\mathit{unknown}$), otherwise $\dleft{(\qENV.\mathit{slice}.y = \qID)}$.
The expression $\qENV.\mathit{slice}.y = \qID$ performs the comparison $y = x$ knowing that~$x$ and~$y$ are not null.
The output of $q$ is of type~$\typeeither$, representing the
three-valued logic of \sql: $\dleft{\mathit{true}}$ is
$\mathit{true}$, $\dleft{\mathit{false}}$ is $\mathit{false}$, and
$\dright{\dunit}$ is $\mathit{unknown}$. The expression
$(\qeither{\qID}{\mathit{false}})$ in the~${\large\sigma}$ converts a boxed three-valued logic value to a Boolean.

\medskip

From \nrae, the translation rewrites the query through a series of
intermediate languages that are successively closer to \imp,
a simple imperative language.  \imp contains variables,
assignments, conditionals, iterations, and calls to external operators
and external runtime functions.
It supports compilation from \sql while remaining easy to translate into various
imperative languages.

The translation of a query in \imp produces a function which is
parameterized by the database instance represented as a record where
each table is a field. The body of the function initializes a
variable \lstjs+ret+ with the result of the query which is returned at the end.
\begin{jslisting}
fun(db) {
  var R = db.R;   var ret;    ...    return ret;
}
\end{jslisting}

In \imp, the code corresponding to $\qmap{ \{ x : \qID.a \} }{R}$ is a
loop that builds a collection \lstjs{tmp0} by iterating over the input
collection provided in \lstjs{R}.
\begin{jslisting}
var tmp0 = array();   for (id0 in R) { tmp0 = push(tmp0, { x : id0.a }) }
\end{jslisting}

Finally, from \imp, we obtain \js code that executes the query via
a straightforward translation.  This is linked to a
\js runtime that implements operations like \lstjs+array+ and
\lstjs+push+.

\section{Main languages}\label{sec:preliminaries}
\sql is a declarative query language built around the famous
\Sql{select from where} statement. While most formal treatments use a
set-theoretic semantics, \sql implementations use a {\it bag}
semantics, i.e., unordered collection in which the same element may
occur multiple times. Most realistic queries use significantly more
complex features, including
\Sql{select from where group by having}
statements to handle aggregation and collection operators such as
$\cup$
(\Sql{union}),
$\cap$
(\Sql{intersect})
and $\setminus$
(\Sql{except})
.
\sql queries have also to account for \null{} values that are used to represent unknown or missing information in tables.
\sql
queries involve {\it predicates}
(\Sql{=,<,...})
{\it functions}
(\Sql{+,-,...})
and {\it aggregates}
(\Sql{sum, max, count,...})
which are often used in conjunction with
\Sql{group by having}. Last, \sql allows for nested expressions, e.g.,
queries inside the \Sql{where} or \Sql{having} clause.

To compile \sql, the three languages that are at the core of the contribution are \sqlalg, \nrae, and \imp.
We present their syntax, data models, and semantics.

\subsection{\sqlalg}
\label{sec:preliminaries:sqlalg}

\sqlalg~\cite{DBLP:conf/cpp/BenzakenC19} is an extension of the relational
algebra to encompass \sql's aggregates, formulas, bag semantics, and
environment handling.
The goal of \sqlalg is to capture the semantics of \sql using the relational algebra as in database textbooks~\cite{DBLP:books/cs/Ullman82,DBLP:books/aw/AbiteboulHV95}, but on a larger fragment than is typically presented.

\sqlalg queries operate on a flat data model.
A database instance is a set of named \emph{relations}~(or \emph{tables}).
Each table is a bag of \emph{tuples} where each element of the tuple is a raw value~(number, string, etc) and can be accessed with its name called \emph{attribute}.
All the tuples in a table have the same set of attributes, the \null value is used to encode a missing attribute.

The syntax of \sqlalg is the following~(the notation $\overline{e}$ indicates a list of expressions $e$).
\[
\small
\begin{array}[t]{cc}
\begin{array}[t]{l@{\hspace*{0.5em}}c@{\hspace*{0.5em}}l}
\algquery & ::= & 
  \mbox{\tt ()} ~~|~~
  \mathit{tbl} \\ && |~~
  \algquery ~(\mathtt{union} ~~|~~ \mathtt{intersect} ~~|~~ \mathtt{except})~ \algquery ~~|~~
  \algquery ~{\large \bowtie}~ \algquery \\ && |~~
  {\large \pi}_{(\overline{e \mathtt{\,as\,} \attribute})}(\algquery) ~~|~~
  {\large \sigma}_{\formula}(\algquery) \\ && |~~
  {\large \gamma}_{(\overline{e \mathtt{\,as\,} \attribute}, \overline{e},\formula)}(\algquery) \\
\end{array}
&
\begin{array}[t]{l@{\hspace*{0.5em}}c@{\hspace*{0.5em}}l}
\formula & ::= &
  \mathtt{true} ~~|~~
  \formula ~(\mathtt{and} ~~|~~ \mathtt{or})~ \formula ~~|~~
  \mathtt{not}~ \formula \\ & & |~~
  p(\overline e) ~~|~~
  p(\overline e, (\mathtt{all} ~~|~~ \mathtt{any})~ {\algquery}) \\ & & |~~
  \overline{e ~\mathtt{as}~ \attribute}~ \mathtt{in}~ {\algquery} ~~|~~
  \mathtt{exists} ~{\algquery}
\\

e   & ::= & c ~~|~~ \attribute ~~|~~ \fn(\overline{e}) ~~|~~ \ag(\overline{e})
\end{array}
\end{array}
\]

 A query $\algquery$ can be a tuple with no attributes ($\mbox{\tt ()}$),
a relation name $\mathit{tbl}$, a set operation on two sub-queries, a
natural join ${\large \bowtie}$,\footnote{A natural join $Q_1~{\large
    \bowtie}~Q_2$ computes the set of all combinations of tuples in
  $Q_1$ and $Q_2$ that are equal on their common attribute names. For
  example, if $Q_1$ computes the bag of tuples
$\{\!|(a:1, b:2), (a:2,b:2), (a:3,b:3)|\!\}$ and $Q_2$ computes the
  bag
$\{\!|(b:1,c:1), (b:2,c:2), (b:2,c:3)|\!\}$, then their natural join
  is the bag
$\{\!|(a:1,b:2,c:2), (a:2,b:2,c:2), (a:1,b:2,c:3), (a:2,b:2,c:3)|\!\}$.
For instance, the tuple $(a:3,b:3)$ from $Q_1$ is discarded since
  there is no tuple in $Q_2$ whose value on the attribute $b$ is $3$,
  whereas the tuple $(a:1, b:2)$ is combined with all the tuples of
  $Q_2$ whose value on the attribute $b$ is $2$.
} a projection (and renaming) ${\large \pi}$, a selection sigma ${\large \sigma}$, or a grouping
${\large \gamma}$. The ${\large \gamma}$ operator extends the standard relational algebra
with the possibility to compute groups and aggregates similarly to a \Sql{select/group by/having} in \sql.
A formula~$\formula$ is an expression returning a Boolean value in the three-valued logic where $p$ is a predicate~(e.g., $<$).
An expression~$e$ can be a constant~$c$ from the set of values $\values$~(currently \dbcert supports intergers, Booleans, and string and also floating point numbers in a separate version), an attribute name~$\attribute$ corresponding to a component of a tuple, or a function call.
There are two classes of functions: (1)~the functions~$\fn$ that combine values~(like $\mathtt{+}$, $\mathtt{-}$, $\mathtt{*}$), and (2)~aggregate functions~$\ag$ that operate over collections~(like~$\mathtt{sum}$, $\mathtt{avg}$, or $\mathtt{min}$).\footnote{In the implementation, expressions with and without aggregate are syntactically stratified.}

\begin{figure}[t]
    {\footnotesize
\[
\setstretch{1.3}
\begin{array}[t]{l@{~}c@{~}l}
\semq{\mbox{\tt ()}}{\env}{i} & = & \{\!| ~ |\!\} \\
\semq{\mathit{tbl}}{\env}{i}  & = & i.\mathit{tbl} \quad \mbox{~if~} \mathit{tbl} \mbox{~is a table}\\
\semq{Q_1 ~{\Large \bowtie}~ Q_2}{\env}{i} & = & \\[\medskipamount]

\multicolumn{3}{l}{\left\{\!\left|\left(\overline{a_n=c_n},\overline{b_k=d_k}\right)
\left|
\begin{array}{l}
(\overline{a_n=c_n}) \in \semq{Q_1}{\env}{i} ~\wedge\\
(\overline{b_k=d_k}) \in \semq{Q_2}{\env}{i}  ~\wedge\\
(\forall~ n,k, ~~ a_n = b_k \Rightarrow c_n = d_k)
\end{array}
\right.\right|\!\right\}} \\
\end{array}
\hspace*{-5.9em}
\begin{array}[t]{l@{~}c@{~}l}
\semq{{\large \pi}_{(\overline{e_n \mathtt{\,as\,} a_n})}(Q)}{\env}{i} & = &
\{\!| (\overline{a_n = \seme{e_n}{(\ell(t),[],[t]) :: \env}}) \mid
t \in \semq{Q}{\env}{i}
|\!\}\\
\semq{{\large \sigma}_{f}(Q)}{\env}{i} & = &
\{\!|t\in\semq{Q}{\env}{i} \mid \semf{f}{(\ell(t),[],[t]) :: \env}{i} = \top|\!\}\\
\multicolumn{3}{l}{
\semq{{\large \gamma}_{(\overline{e_k \mathtt{\,as\,} a_k},\overline{e_n},f) }(Q)}{\env}{i} = } \\[\medskipamount]
\multicolumn{3}{@{\hspace{2cm}}l}{\left\{\!\left| \overline{(a_k=\seme{e_k}{(\ell(T),\overline{e_n},T)::\env})} | T\in\mathbb{F}_3 \right|\!\right\}} \\
\multicolumn{3}{@{\hspace{3cm}}l}{\mbox{and~} \mathbb{F}_2 \mbox{~is a partition of $\semq{Q}{\env}{i}$ according to $\overline{e_n}$} }\\
\multicolumn{3}{@{\hspace{3cm}}l}{\mbox{and~} \mathbb{F}_3 = \left\{\!\left|T \in {\mathbb F}_2  \left| \semf{f}{(\ell(T),\overline{e_n},T)::\env}{i} = \top\right.\right|\!\right\}} \\
\end{array}
\]
\smallskip
\[
\setstretch{1.3}
\begin{array}[t]{l@{~}c@{~}l}
  \semf{f_1 \mathtt{ ~and~} f_2}{\env}{i} & = & \semf{f_1}{\env}{i} \wedge_3 \semf{f_2}{\env}{i}\\
  \semf{f_1 \mathtt{ ~or~} f_2}{\env}{i} & = & \semf{f_1}{\env}{i} \vee_3 \semf{f_2}{\env}{i}\\
\semf{p(\overline{e_n})}{\env}{i} & = & p(\overline{\sema{e_n}{\env}})\\
  \end{array}
  \quad
  \begin{array}[t]{l@{~}c@{~}l}
  \semf{p(\overline{e_n}, \mathtt{~all~} q)}{\env}{i} & = & \mathit{true}
  ~~{\mbox{iff~} \semf{p(\overline{e_n},t)}{\env}{i} = \mathit{true} \mbox{~for all~} t \in \semq{q}{\env}{i}}\\
\semf{\mathtt{exists~} q}{\env}{i} & = & \mathit{true}
  ~~\mbox{iff~} \semq{q}{\env}{i} \mbox{~is not empty}
  \end{array}
\]
\medskip
\[
\setstretch{1.3}
\begin{array}[t]{l@{~}c@{~}l}
  \seme{c}{\env} & = & $c$
\end{array}
\quad
\begin{array}[t]{l@{~}c@{~}l}
\seme{\fn(\overline{e})}{\env} & = & \fn(\overline{\seme{e}{\env}})
\end{array}
\quad
\begin{array}[t]{l@{~}c@{~}l}
  \seme{a}{(A, G, T) :: \env} & = &
  \left\lbrace
  \begin{array}[c]{@{~}ll}
   T.a  & \mbox{if~} a \in A\\
  \seme{a}{\env} & \mbox{if~} a \notin A
  \end{array}
  \right.
\end{array}
\]

}

   \caption{Semantics of \sqlalg (excerpt).}
\label{fig:sqlalg-semantics-summary}
\end{figure}

The semantics of \sqlalg is defined in
\Cref{fig:sqlalg-semantics-summary}~(the full definition is given in
\ifextended \Cref{fig:sqlalg-semantics,fig:sqlalg-formula-semantics,fig:algsem} of \Cref{sec:semantics-app}\fi{}).
The semantics function of each syntactic category is annotated with the syntactic category of the term~($Q$ for queries, $f$ for formulas, and $e$ for expressions).
The semantics $\semq{Q}{\env}{i}$ of a query~$Q$ evaluated in an environment~$\env$ on a database instance $i$ defines a bag.
The instance~$i$ associates the data to each table.
The environment~$\env$ defines the local evaluation context of the
query, and is a major subtlety in the semantics of \sql that we now
detail.

An environment \env = $[S_n;...;S_1]$ has a stack structure reflecting the current nesting level of the query.
Each level of the stack is a \emph{slice} $S = (A, G, T)$ where
$A$ is the set of attributes defined at the slice level,
$G$ is the list of grouping expressions in the case of a ${\large \gamma}$, and
$T$ is the current tuple (or list of tuples for a ${\large \gamma}$) the query is evaluated against.
We use the notation $\env = S::\env'$ to access the top of the stack and
$A(S)$, $G(S)$, $T(S)$ to access to the different elements of a slice.

The rules for the
projection~(${\large \pi}_{(\overline{e_n \mathtt{\,as\,}
    \attribute_n})}(\algquery)$) and for the grouping
operator~(${{\large \gamma}_{(\overline{e_k \mathtt{\,as\,}
      a_k},\overline{e_n},f) }(Q)}$) in \Cref{fig:sqlalg-semantics-summary}
illustrate the construction of the environment. Projection
builds a bag by iterating on each tuple~$t$ in the result of the
evaluation of~$\algquery$. For each~$t$, it creates a new tuple with
attributes~$\overline{\attribute_n}$ of value $\overline{e_n}$. Each
expression $e_n$ is evaluated in an environment
$\env' = (\ell(t), [], [t])::\env$ where the function~$\ell$ extracts
the attribute names of~$t$. The grouping operator, in contrast, first
agglomerates tuples as per some grouping expressions
$\overline{e_n}$, then filters out some of the groups using the
predicate~$f$, and finally computes the resulting expressions
$\overline{e_k}$ on groups. Thus, the predicate and the resulting
expressions are evaluated in an environment extended with the groups
computed in $\mathbb{F}_3$.

\begin{example}
\label{sec:preliminaries:sqlalg:example}
Consider the query \Sql{Q1} from \Cref{subsec:examples} in \sql and its
corresponding \sqlalg expression:\footnote{For simplicity, we omit
  renamings that would have been added by \sqlcoq.}
\begin{center}
\Sql{select a1 from t1 group by a1 having exists}
\Sql{                  (select a2 from t2 group by a2 having sum(1+0*b2) = 2);}
\end{center}
$$
{\large\gamma}_{((a_1\mathtt{\,as\,}a_1),\, a_1,\, \mathtt{exists}({\large\gamma}_{\scriptstyle((a_2\mathtt{\,as\,}a_2),\, a_2,\, \mathtt{sum}(1+0*b_2) = 2)}(t_2)))}(t_1)
$$

Following the semantics of \Cref{fig:sqlalg-semantics-summary} with
$t_1 = \{\!| (a_1: 1, b_1: 1), (a_1: 1, b_1: 2), (a_1: 2, b_1: 3), (a_1: 3, b_1: 1), (a_1: 3, b_1: 2), (a_1: 3, b_1: 3) |\!\}$,
the ${\large\gamma}$ operator first creates the partition of $t_1$
according to the value of the attribute $a_1$:
$$\begin{array}{l}
\mathbb{F}_2 = [ T_1, T_2, T_3] \text{ with } T_1 = \{\!| (a_1: 1, b_1: 1), (a_1: 1, b_1: 2) |\!\} \text{, } T_2 = \{\!| (a_1: 2, b_1: 3) |\!\}\\

\qquad\qquad\,\,\,\qquad \text{ and } T_3 = \{\!| (a_1: 3, b_1: 1), (a_1: 3, b_1: 2), (a_1: 3, b_1: 3) |\!\}
\end{array}
$$
The formula $\mathtt{exists}(...)$ is evaluated on each group $T$ in $\mathbb{F}_2$ in an environment~$\env = ([a_1, b_1], a_1, T)$.
If $t_2 = \{\!| (a_2: 7, b_2: 7), (a_2: 7, b_2: 8) |\!\}$, similarly,
the nested query is a ${\large\gamma}$ operator that creates the partition of $t_2$
according to the value of the attribute $a_2$:
$$
\mathbb{F}'_2 = [ \{\!| (a_2: 7, b_2: 7), (a_2: 7, b_2: 8) |\!\} ]
$$
So the formula
$\mathtt{sum}(1+0*b_2) = 2$ is evaluated in an
environment~$\env' = [([a_2, b_2], a_2, \{\!| (a_2: 7, b_2: 7), (a_2: 7, b_2: 8) |\!\}), ([a_1, b_1], a_1, T)]$
for each $T \in \mathbb{F}_2$. The evaluation of this formula will evaluate $b_2$ (even
if it is multiplied by $0$): $b_2$ appears twice in the first slice of
the environment, so no matter~$T$, the expression~$1+0*b_2$ is summed twice and
$\mathtt{sum}(1+0*b_2) = 2$ is true for any $T$. This is why the entire
query \Sql{Q1} returns the three grouping values for $a_1$: $1$, $2$ and
$3$, thus the bag
${\{\!| (a_1: 1), (a_1: 2), (a_1: 3) |\!\}}$.

For the query \Sql{Q2}, the same reasoning holds, except that the inner
formula is $\mathtt{sum}(1+0*b_1) = 2$. This time, the evaluation of
this formula will evaluate $b_1$, which appears in the second slice.
Thus, $1+0*b_1$ is summed twice for $T_1$, once for $T_2$ and three
times for $T_3$, which makes
$\mathtt{sum}(1+0*b_1) = 2$ valid only for $T_1$. The result of the query
is thus only the grouping value
$a_1 = 1$, thus the bag
${\{\!| (a_1: 1) |\!\}}$. \hfill $\blacksquare$
\end{example}

The semantics of formulas uses a three-valued logic where Boolean values can be $\mathit{true}$, $\mathit{false}$, or $\mathit{unknown}$.
The value $\mathit{unknown}$ is introduced by predicates on the \null value.
For example, the result of $1 > \text{\null}$ is $\mathit{unknown}$.
The operators in this logic are noted $\wedge_3$, $\vee_3$, and
$\neg{}_3$. They provide the maximum information, so for example
$\mathit{true} \vee_3 \mathit{unknown} = \mathit{true}$ and
$\mathit{false} \vee_3 \mathit{unknown} = \mathit{unknown}$.

The rule for accessing an attribute~$\attribute$ in \Cref{fig:sqlalg-semantics-summary}
looks up a value in the environment, which is traversed from the top of the stack until the
attribute is found, as we have seen in the example. The same process is used
for aggregates $\ag{}$, but a group is searched for in the environment
instead of a single attribute.
Depending on the slice, $T$ can be either a tuple~(if it is introduced by a anything but grouping) or a list~(if introduced by grouping).
The well formedness of the query guaranties that the access to an attribute ($T.\attribute$) can only occur on a tuple and not a list.

\subsection{\nrae}
\label{sec:preliminaries:nrae}

\nrae~\cite{DBLP:conf/sigmod/AuerbachHMSS17} is an extension of
Nested Relational Algebra~\cite{DBLP:conf/dbpl/CluetM93}, designed for optimizations.
For example, there is no stratification between expressions, formulas, and queries, which allows cross level rewriting.
Some optimizations strategies already exist for this language~\cite{DBLP:conf/dbpl/CluetM93,claussen1997optimizing,DBLP:journals/tkde/ClaussenKMPS00,querycompilers}.

As suggested by the name, \nrae supports nested data:

\begin{gram}
  \mbox{}~d & ::= & c
  \mid \{\overline{A_n: d_n}\}
  \mid [\overline{d_n}]
  \mid \dleft{d}
  \mid \dright{d}
\end{gram}

A value is either an atom (a constant), a record, a bag, or a value of type {\tt either}~(a value with a constructor {\dleft{}} or {\dright{}}).
A record is a mapping from a finite set of attributes to values.
A large set of atoms is supported including numbers, strings, and Booleans.
Values of type {\tt either} are used to encode \sqlalg's typed null values and three-valued logic~(\Cref{sec:sqlatonrae:translation:exptran}).

The syntax of the language is the following:
\begin{gram}
  \mbox{}~q & ::= & d
     \mid \qID
     \mid \opunop\, q \mid q_1 \opbinop q_2
     \mid    q_2 \circ q_1 \mid    \qmap{q_2}{q_1}   \mid  \qselect{q_2}{q_1}  \mid    q_1 \times q_2      \mid    q_1~\qor~q_2        \mid   \qeither{q_1}{q_2}        \\ &&
     \mid \qENV             \mid   q_2 \circ^{e} q_1  \mid   \qmapenv{q}       \mid \qgroupby{g}{\overline{a}}{q}
   \end{gram}

A query $d$ returns the value $d$.
The query $\qID$ returns the data $d$ it is evaluated against.
The queries $\opunop\, q$ and $q_1 \opbinop q_2$ represent the application of unary operators~(like negation, field access, building a singleton collection) and binary operators~(like union of bags, record concatenation).

The query composition $q_2 \circ q_1$ illustrates the combinatorial
nature of the semantics. It first evaluates $q_1$ on the input data,
then uses the result of the evaluation to evaluate $q_2$.  The query
$\qmap{q_2}{q_1}$ evaluates the query $q_2$ on each element of the bag
returned by the evaluation of~$q_1$.  The operators
$\qselect{q_2}{q_1}$ and $q_1 \times q_2$ are, respectively,
selection and Cartesian product.  The semantics of product use the
$\oplus$ binary operator, which performs record concatenation.  For
example, the expression $\{a:true\}\oplus \{b:3\}$ evaluates to
$\{a:true, b:3\}$.

The operators $q_1~\qor~q_2$ and $\qeither{q_1}{q_2}$ are control
structures. The query $q_1~\qor~q_2$ checks the result of running $q_1$
on the input data.  If it is not an empty bag it returns it, otherwise it evaluates $q_2$ on the input data.
The query $\qeither{q_1}{q_2}$ matches the input data with ${\dleft{d}}$ and ${\dright{d}}$ and executes either $q_1$ or $q_2$ on the data~$d$ as appropriate.

The queries $\qENV$, $q_2 \circ^{e} q_1$, and $\qmapenv{q}$ 
manipulate the local environment~$\envnraenv$.
$\qENV$ returns the environment~$\envnraenv$, $q_2 \circ^{e} q_1$ updates it, and $\qmapenv{q}$ iterates over it.

\nrae also provides a $\qgroupby{g}{\overline{a}}{q}$ construct that
evaluates $q$ and groups the result using the values of the fields
$\overline{a}$ as keys. The result is a collection of records made of the
key and a field $g$ containing the associated group. For example,
$\qgroupby{g}{x}{d}$ where $d = [ \{ x: 1, y: 1 \}, \{ x: 1, y: 2 \}, \{ x: 2, y: 3
  \} ]$ returns
$[ \{ x: 1, g: [ \{ x:1, y: 1 \}, \{ x: 1, y: 2 \} ] \}, \{ x: 2, g: [
\{ x: 2, y: 3 \} ] \} ]$.

This construct can be defined using simpler constructs of \nrae as
follows:
\[
\begin{array}{lcl}
\qgroupby{g}{\overline{a}}{q} & = &
 \chi_{\left\langle{\qID \oplus \{ g : \qselect{\qENV.\mathit{key} = \pi[\overline{a}](\qID) }{\qENV.\mathit{input}} \circ^{e} (\{\mathit{key}:\qID\} \oplus \qENV)}\right\rangle}\\
  &&\qquad\left(\kkeyword{distinct}(\qmap{\pi[\overline{a}](\qID)}{\qENV.\mathit{input}})\right) \\
  &&\circ^{e} \{ \mathit{input} : q \}
  \end{array}
\]

It is easiest to understand how this definition works by proceeding
backwards.  The last line creates a record with a single label named
$\mathit{input}$ that contains the result of evaluating
$q$. The~$\circ^{e}$ expression causes this record to be used as the
environment when evaluating the preceding lines.

The middle line constructs the set of distinct keys by iterating over
$\mathit{input}$. It uses two operators: $\kkeyword{distinct}(d)$
which takes a bag~$d$ and removes the duplicates, and record
projection $\pi[\overline{a}](d)$ which takes a record~$d$ and returns
the same record with only the specified labels $\overline{a}$.
Since $\mathit{input}$ is stored in the current environment~(thanks to
the third line), it can be accessed by $\qENV.\mathit{input}$.  Using
map~($\chi$) and project~($\pi$), we extract the keys
from $\mathit{input}$ and use $\kkeyword{distinct}$ to ensure they are unique.

The last line maps over the set of distinct keys.  For each one, we
are building a record containing the key and a field~$g$ constructed
as follow.
We first extend our environment with an additional field
$\mathit{key}$ containing the current
key~($\{\mathit{key}:\qID\} \oplus \qENV$).
Then, we select the records in $\mathit{input}$ matching the
key~($\qselect{\qENV.\mathit{key} = \pi[\overline{a}](\qID)}{\qENV.\mathit{input}}$).

\begin{figure*}[t]
    \begin{minipage}[l]{\textwidth}
{\footnotesize
  \begin{gather*}
    \infer[Constant]
    {\envnraenv \vdash d_0 \qapp\ d \Downarrowa d_0}
    {}
    \qquad
    \infer[ID]
    {\envnraenv \vdash \qID \qapp\ d \Downarrowa d}
    {}
    \qquad
    \infer[Comp]
    {\envnraenv \vdash q_2 \circ q_1 \qapp\ d_0 \Downarrowa d_2}
    {\envnraenv \vdash q_1 \qapp\ d_0 \Downarrowa d_1
     \quad
     \envnraenv \vdash q_2 \qapp\ d_1 \Downarrowa d_2}
    \\[\jot]
\infer[Env]
      {\envnraenv \vdash \qENV \qapp\ d \Downarrowa \envnraenv}
      {}
    \qquad
    \infer[\mbox{Comp$^e$}]
      {\envnraenv_1 \vdash q_2 \circ^e q_1 \qapp\ d_1 \Downarrowa d_2}
      {\envnraenv_1 \vdash q_1 \qapp\ d_1 \Downarrowa \envnraenv_2
       \quad
       \envnraenv_2 \vdash q_2 \qapp\ d_1 \Downarrowa d_2}
    \\[\jot]
\infer[\mbox{Either$_{\dleft{}}$}]
      {\envnraenv \vdash \qeither{q_1}{q_2}\qapp {\dleft{}}\ d \Downarrowa d_1}
      {\envnraenv \vdash q_1\qapp d \Downarrowa d_1}
    \qquad
    \infer[\mbox{Either$_{\dright{}}$}]
      {\envnraenv \vdash \qeither{q_1}{q_2}\qapp {\dright{}}\ d \Downarrowa d_2}
      {\envnraenv \vdash q_2\qapp d \Downarrowa d_2}
  \end{gather*}
}
\end{minipage}

   \caption{Semantics of \NRAEnv~(excerpt).}
  \label{fig:nra-semantics-summary}
\end{figure*}

The formal semantics of \nrae is defined by a judgment ${\envnraenv \vdash q \rapp d \Downarrowa d'}$ which means that a query $q$ evaluated in a local environment $\envnraenv$ against input data $d$ produces a value~$d'$ where the environment~$\envnraenv$ can be any \nrae data~(\eg a record or a collection).
A few rules are given in \Cref{fig:nra-semantics-summary}. The complete
semantics can be found in \ifextended \Cref{fig:nra-semantics} of \Cref{sec:semantics-app}\fi{}.

Compared to the original algebra, we have replaced the operator $q_1 |\!| q_2$ which was testing if the input was the empty collection by the operator $\qeither{q_1}{q_2}$ which corresponds to the pattern matching on the values $\dleftOp$ and $\drightOp$.

\subsection{Imp}
\label{sec:preliminaries:imp}

The goal of \imp, the final language we define, is to be close to the targeted runtime.
It is parameterized by a data model~(the constant values), the
built-in operators~(like addition), and the library functions~(the
runtime needed to execute the program).
\imp can be instantiated into a subset of most imperative languages.
The syntax of \imp is the following:
\[
  \begin{array}{lcl}
    e & ::= &
              c
              ~|~ x
              ~|~ \mathit{op}(e)
              ~|~ f(e)
    \\
    s & ::= &
              \impblock{ \mathit{decl}^* s^* }
              ~|~ \impassign{x}{e}
              ~|~  \impforrange{x}{e}{s}
              ~|~ \impif{e}{s}{s}
    \\
    \mathit{decl} & ::= &
                  \impvar{x}
              ~|~ \impvarinit{x}{e}
    \\
    q & ::= &
              \impfun{x}{s}{y}
  \end{array}
\]
A query $q$ is a function that takes an argument~$x$ as the input data.
Its body is an imperative statement~$s$ that must define the value of the returned variable~$y$.
Statements can be assignments, loops over a collection, conditionals and blocks.
A block can contain a list of variable declarations, that can be initialized or not, followed by a sequence of statements.
Finally, expressions are constants~($c$), variables~($x$), operator applications~($\mathit{op}(e)$), and runtime operator calls~($f(e)$).

The semantics $\impsem{e}{\envimp}$ evaluates in an environment $\envimp$ an expression $e$ into a value~$c$ and $\impsem{s}{\envimp}$ evaluates statement~$s$ into a new environment~$\envimp'$. 
The definition is standard and given in \ifextended \Cref{fig:imp-sem} of \Cref{sec:semantics-app}\fi{}.
The only particularity is that an instantiation of \imp must provide a
semantics~$\impdatasem{.}{}$ for the parameterized operator and
library functions.
The instantiation must also provide two functions $\tobool{c}$ and
$\tolist{c}$ that reify values of the language respectively into a
Boolean and into a list.  Assuming that all the instantiated operators are terminating, all Imp programs are terminating.
For example, the semantics of the conditional~($\impsem{\impif{e}{s_1}{s_2}}{\envimp}$) has to interpret the result of the evaluation of~$e$ as a Boolean.
The two expressions~$\mathit{op}(e)$ and~$f(e)$ have the same
semantics. They are separate in \imp to distinguish functions that are
compiled into a built-in operator in the target language, with
functions that are compiled into a runtime library function. For
example, the addition between two integers is compiled into a runtime
library function in \js, but it could be compiled into a built-in
operator if we target another language that supports integers.

\medskip

\newcommand{\ejsnull}{\ensuremath{\texttt{null}}}
\newcommand{\ejsobj}[1]{\ensuremath{\texttt{\{}\ {#1}\ \texttt{\}}}}
\newcommand{\ejsarray}[1]{\ensuremath{\texttt{[}\ {#1}\ \texttt{]}}}

As our target is \js code, we instantiate the Imp data model with EJson, an extended JSON with integers and functional arrays~(the $\ejsnull$ of \js does not have the same semantics of the \null of \sql):
$$
\begin{array}[t]{lcl}
  c & ::= & \mbox{string~val} ~|~ \mbox{number~val} ~|~ \mbox{bool~val} ~|~ \ejsnull ~|~ \ejsobj{\overline{ l_n : c_n }} ~|~ \mbox{integer~val} ~|~ \ejsarray{\overline{c_n}}
\end{array}
$$

The operators are those of the host language. For example,
the operator~\lstinline{*} corresponds to multiplication on \js
numbers~(IEEE754 floating point numbers).
Supporting \sql requires Boolean
 arithmetic and string operators, as well as comparisons and access to the fields of an object.

Finally, the instantiation of \imp also comes with runtime functions
that need to be implemented in \js. Examples of such functions are
operations on integers and functional arrays.

\section{From \sqlalg to \nrae}\label{sec:sqlatonrae}
This section presents the translation between \sqlalg and \nrae.
The key aspects are the following:
\begin{enumerate}
\item\label{sec:sqlatonrae:item:challenge1} encoding \null values and three-valued logic, and the operations
  on them;
\item\label{sec:sqlatonrae:item:challenge2} reflecting the environments and how they are handled.
\end{enumerate}

The first challenge is that \nrae does not support three-valued logic connectives.  To address this, we 
encode these connectives, as discussed in \Cref{sec:sqlatonrae:datamodel}.
The second challenge arises from
the subtle handling of environments presented in \Cref{sec:preliminaries:sqlalg}.
We address this by making the \sqlalg environment explicit in the generated \nrae expression~(\Cref{sec:sqlatonrae:translation:envtran,sec:sqlatonrae:translation:queries}).
These challenges in the translation, presented in this section, are also
reflected in the proof, as explained in~\Cref{sec:related:evaluation:challenges}.

\sqlalg is a stratified language with multiple syntactic categories
(queries, formula, and expressions), whereas \nrae is an expression
language. Similarly the \sql data are stratified
but not the \nrae ones. This complicates the translation and proofs.
Notationally, we index each translation function with the syntactic category of its argument ($Q$ for queries, $f$ for formula, ...).

Before detailing the translation from \sqlalg to \nrae, we state the (verified) correctness theorems.

\subsection{Correctness}
\label{sec:sqlatonrae:correctness}

Correctness of the translation applies only to \textit{meaningful}
\sqlalg queries:
queries that are well typed and well formed~\cite{DBLP:conf/cpp/BenzakenC19}.
For simplicity, we elide this assumption in the following theorems.

\Cref{thm:top} is the main theorem. It states
the correctness of the translation from \sqlalg to \nrae: for any query
$Q$, the evaluation of its translation to \nrae ($\trq{}{Q}$) on the translated instance~$\tri{i}$ is equal to
the translation of the result of its evaluation ($\trbag{\semq{Q}{}{i}}$).

\begin{theorem}\label{thm:top}
  $\forall Q~i,
  ( \vdash \trq{}{Q} \qapp \tri{i} \Downarrowa
  \trbag{\semq{Q}{}{i}})\QED$
\end{theorem}

\Cref{thm:glob} generalizes \Cref{thm:top} to any environment, which
is needed for sub-queries. As we will detail
in~\Cref{sec:sqlatonrae:translation:envtran}, the \sqlalg environment
$\env$ can be seen has having a
statically determinable part ($\taenv{\env}$, containing the groups and
attribute names but not the data) that is used at compile-time,
while its dynamic part actually
containing the tuples of the sub-queries ($\tenv{\env}$) will be
available at runtime. The generalized theorem thus states
that for any query $Q$ and any environment $\env$, the evaluation of
the translation of $Q$ in the dynamic environment $\tenv{\env}$ is
equal to the encoding (using the static environment $\taenv{\env}$) of
the evaluation of $Q$ into \nrae data model.

\begin{theorem}\label{thm:glob}
  $\forall Q~\env~i,
   ( \tenv{\env} \vdash \trq{\taenv{\env}}{Q} \qapp \tri{i} \Downarrowa
   \trbag{\semq{Q}{\env}{i}})\QED$
\end{theorem}

\begin{figure*}[t]
\begin{tabular}{cc}

\tikzstyle{box}=[minimum width=1.5cm,
minimum height=.75cm,rounded corners=2pt,
draw,fill=white,text=black,font=\normalfont]

\tikzstyle{txt}=[minimum width=1.5cm,
minimum height=.75cm,rounded corners=2pt,
fill=white,text=black,font=\normalfont]

\tikzstyle{arrw}=[->,>=stealth,rounded corners=5pt,thick]
\tikzstyle{eval}=[->,shorten <=1pt,snake=snake,line before snake=2pt]

\begin{minipage}[c]{.45\linewidth}
{\Large
  \centering
  \resizebox{\textwidth}{!}{\begin{tikzpicture}
      \node[txt,
      label=
      {[align=center]
        above:{\sqlalg}}] at (0,-1) (q)
      {\huge  $Q$};\node[txt,
      label=
      {[align=center]
        above: {\nrae}},
      right= 8 of q]
      (n) {\huge  $q$};\node[txt,
      label=
      {[align=center]
        below: {Bag of tuples}},
      below= 3 of q]
      (b) {\huge $b$};\node[txt,
      label=
      {[align=center]
        below: {\data collection }}
      ,below = 3 of n]
      (d) {\huge $d$};\draw[eval]   ($(q.south)$) -- ($(b.north)$)
      node[midway,left]  (labq){\larger $\semq{Q}{\env}{i}$}
      node[midway,sloped,above]
      {Evaluation};\draw[eval]   ($(n.south)$) -- ($(d.north)$)
      node[midway,left] (labn)
      {\larger
        $( \tenv{\env}
        \vdash q \qapp \tri{i}
\Downarrowa
        d) $}
       node[midway,sloped,above]{Evaluation};\draw[arrw] (q.east) --  ($(n.west)$)
      node[midway,above] (labqn)
      {\larger $\trq{\taenv{\env}}{Q}$}
      node[midway,below]  {Translation};\draw[arrw] ($(b.east)+(0,.1)$) --  ($(d.west)+(0,.1)$)
      node[midway,above]  {Encoding: \larger$\trbag{b} \equiv_{d} d$};\end{tikzpicture}
    }}
\end{minipage}

  &

\tikzstyle{box}=[minimum width=1.5cm,
minimum height=.5cm,rounded corners=2pt,
draw,fill=white,text=black,font=\normalfont]

\tikzstyle{box2}=[minimum width=1.5cm,
minimum height=3.cm,rounded corners=2pt,
draw,fill=white,text=black,font=\normalfont]

\tikzstyle{txt}=[minimum width=1.5cm,
minimum height=.75cm,rounded corners=2pt,
fill=white,text=black,font=\normalfont]

\tikzstyle{arrw}=[->,>=stealth,rounded corners=5pt,thick]
\tikzstyle{eval}=[->,shorten <=1pt,snake=snake,line before snake=2pt]

\begin{minipage}[c]{.4\linewidth}
{\Large
  \centering
  \resizebox{\textwidth}{!}{\begin{tikzpicture}
      \node[box,
      label=
      {[align=center]
        above:{}}] at (0,-1) (sql)
      {\Large  \sqlalg};\node[box,
      label=
      {[align=center]
        above:{}},above left= .5 and .4 of sql]  (form)
      {\Large  Formulas};\node[box,
      label=
      {[align=center]
        above:{}},above left= .5 and .1 of form]  (exp)
      {\Large  Expressions};\node[box2,
      label=
      {[align=center]
        above: {}},
      above right= -0.7 and  6 of sql]
      (nra) {\Large  \nrae};\node[box,
      label=
      {[align=center]
        below: {}},
      below= 1.95 of sql]
      (bag) {\Large Bag};\node[box2,
      label=
      {[align=center]
        below: {}}
      ,below = 1.9 of nra]
      (data) {\Large \data};\node[box,
      label=
      {[align=center]
        above:{}},below = 4.25 of form]  (log)
      {\Large  3v-logic};\node[box,
      label=
      {[align=center]
        above:{}},below = 6.5 of exp]  (val)
      {\Large  Values};

      \draw[eval]   ($(sql.south)$) -- ($(bag.north)$)
      node[midway,left]  (labq){}
      node[midway,sloped,below,rotate=180]
      {\footnotesize Evaluation};\draw[eval]   ($(nra.south)$) -- ($(data.north)$)      
      node[midway,left] (labn)
      {}
      node[midway,sloped,below,rotate=180]
      {\footnotesize Evaluation};\draw[arrw] (sql.east) --  ($(nra.west)-(0,1.15)$)
      node[midway,sloped,above]
      (labqn)
      {}
      node[midway,sloped,below,rotate=360]
      {$\trq{}{\_}$};\draw[arrw] (bag.east) --  ($(data.west)+(0,1.1)$)
      node[midway,sloped,above]
      {}
      node[midway,sloped,below,rotate=360]
      {$\trbag{\_}$};\draw[arrw] (log.east) --  ($(data.west)-(0,0)$)
      node[midway,sloped,above]
      {}
      node[midway,sloped,below,rotate=360]
      {$\trB{\_}$};\draw[arrw] (val.east) --  ($(data.west)-(0,1.1)$)
      node[midway,sloped,above]
      {}
      node[midway,sloped,below,rotate=360]
      {$\trval{\_}$};

      \draw[arrw] (form.east) --  ($(nra.west)$)
      node[midway,sloped,above]      
      {}
      node[midway,sloped,below,rotate=360]
      {$\trf{}{\_}$};

      \draw[arrw] (exp.east) --  ($(nra.west)+(0,1.1)$)
      node[midway,sloped,above]      
      {}
      node[midway,sloped,below,rotate=360]
      {$\tre{}{\_}$};

      \draw[eval]   ($(form.south)$) -- ($(log.north)$)
      node[midway,left]  {}
      node[midway,sloped,below,rotate=180]
      {\footnotesize Evaluation};

      \draw[eval]   ($(exp.south)$) -- ($(val.north)$)
      node[midway,left]  {}
      node[midway,sloped,below,rotate=180]
      {\footnotesize Evaluation};

    \end{tikzpicture}
    }}
\end{minipage}

  \\
\end{tabular}
\caption{\sqlalg to \nrae correctness diagram: top-level (left,
  \Cref{thm:top,thm:glob}) and internally (right,
  \Cref{thm:glob,thm:form-sem,thm:exp-sem}).}
\label{fig:trans1}
\label{fig:trans:mult}
\end{figure*}

An alternative view to \Cref{thm:glob} is given in
\Cref{fig:trans1} (left).

The translation of queries~($\Tau^{Q}$) relies on the translation of formulas~($\Tau^{f}$) and expressions~($\Tau^{e}$).
\Cref{thm:form-sem,thm:exp-sem} establish semantics preservation of these translations.
To compare evaluations of \sql formulas and expressions and their corresponding \nrae expressions, \sql Booleans and values are translated into \nrae data~($\Tau^{\mathsf{B}}$, and$\Tau^{\mathsf{val}}$).
\Cref{fig:trans:mult} (right) presents these theorems graphically.

\begin{theorem}\label{thm:form-sem}
$\forall f~\env~i, ( \tenv{\env} \vdash \trf{\taenv{\env}}{f} \rapp \tri{i}
\Downarrowa \trB{\semb{f}{\env}{i}})\QED$
\end{theorem}

\begin{theorem}\label{thm:exp-sem}
  $\forall e~\env, ( \tenv{\env} \vdash \tre{\taenv{\env}}{e}
    \rapp []
    \Downarrowa \trval{\seme{e}{\env}})\QED$
\end{theorem}

\subsection{Translation of the data model}
\label{sec:sqlatonrae:datamodel}

Database instances are the contents of the relations~(tables).
Since we consider only \Sql{select} queries, the database instance is constant during evaluation.
For \sqlalg, it is defined as a record where fields are labeled with table names and values are bags of tuples.
For \nrae, it is defined as a record where values can be any \data.
The function~$\Tau^{\mathsf{i}}$ translates database instances, mapping relation names to \nrae's labels~($\Tau^{\mathsf{tab}}$).
Each \sqlalg tuple is translated into a \nrae record where the name of each attribute is mapped to a label~($\Tau^{\mathsf{att}}$) and each value is translated into a \nrae data~($\Tau^{\mathsf{val}}$).

Value translation needs to handle \null.  This
is done using an option type for nullable values. Following
\nrae convention, this is represented by boxing each value in a data of type $\typeeither$, which can be $\dleftOp$ or $\drightOp$.
A non-null value~$v$ is represented as $\dleft{v}$ and a null value is represented as $\dright{\dunit}$.

To encode the three-valued logic in \nrae, we also use the $\typeeither$ type:
$$
\mathit{true}_{3} = {\dleft{\true}}
\qquad
\mathit{false}_{3} = {\dleft{\false}}
\qquad
\mathit{unknown} = {\dright{\dunit}}
$$
We define the operators $\neg_B,\wedge_B,\vee_B$ and ${\tt is\_true}_B$ as \nrae expressions.
For example, the $\neg_B$ operator is implemented as follows:
\[
\begin{array}{lcl}
  \neg_B~q
  &=&
      {(\qeither{\dleft{(\neg~\qID)}}{\dright{\dunit}})}
    \circ q
\end{array}
\]
This expression uses the~$\circ$ operator to first evaluate $q$ to a
data~$d$ and then give $d$ as input to
$(\qeither{\dleft{\neg~\qID}}{\dright{\dunit}})$.  Then the matching
operator~($\qeither{q_1}{q_2}$) returns $\dleft{\neg~b}$ if
$d = \dleft{b}$ otherwise it returns $\dright{\dunit}$.
So $\neg_B~q$ has the expected behavior of returning \texttt{unknown}
if $q$ is \texttt{unknown} and the negation of the Boolean otherwise.

\subsubsection{Floating Points and Bags: an Inconsistency}
\label{sec:sqlatonrae:datamodel:float}

DBcert supports Boolean values, integers, and strings.
The SQL specification also supports floating point operations.
Unfortunately, however, our initial attempts at supporting them
revealed that the aggregate operators \Sql{sum} and \Sql{avg} are not
compatible with the set and bag semantics of \sql despite both being
mainstream features of Relational Database Management Systems~(RDBMSs).

Indeed, the specification requires for the aggregates \Sql{sum} and
\Sql{avg} that addition is
associative (and commutative), as aggregates are operating over
(unordered) bags; but floating point addition is not associative. This
issue is a fundamental problem with using floating point aggregate
operations over unordered bags, and is a~(mostly ignored) problem in
real Relational Database Management Systems.

Our base compiler avoids this inconsistency by eliding support for
floating point.  However, to target queries based on realistic JSON
databases, supporting floating point operations is important.

There are a number of possible solutions to this issue.  We
could acknowledge that floating point addition is indeed
non-associative, and model that in the semantics.  If we keep a bag
semantics, this would change the semantics to be non-deterministic.
If we change from using a bag semantics, we have a different
infelicity to traditional semantics that would significantly inhibit
query optimization.

Alternatively, we can avoid the non-associativity of
floating point by using a slight-of-hand often
employed in theorem proving: changing from modelling floating point
numbers to modelling real numbers.  This would regain associativity,
but be unfaithful to our extracted implementation.

Another option is to keep modelling floating point numbers, but
pretend that the \Sql{sum} and \Sql{avg} aggregate operators are
associative.  This keeps the model simple, but
introduces (false) axioms, that need to be carefully isolated so they
do not infect the rest of the verification effort.

We created a variant of our compiler that proceeds along the lines of
the last option.  We extended \sqlalg with double precision floating
point values, using Coq's native floats.  We also extended the
functions $\fn$ with arithmetic and Boolean operations on these
values, and aggregate operators $\ag$ with \Sql{sum}, \Sql{max}, and
\Sql{avg}.  This pragmatic approach continues to model and reason
about floating point numbers, while pretending that addition is
associative and commutative by assuming these properties as axioms.
While these axioms are technically unsound, we took great care to
isolate their usage to these proofs of the floating point aggregates.
Note that three other axioms about floating point numbers are also
assumed, but these are all valid: associativity and commutativity of
floating point maximum, and a specification for injecting positive
integers into floats.  These are specified as axioms since they
characterize functions not implemented in Coq and only realized during
extraction, however they are believed to be sound.

In addition to the care taken to isolate the use of these unsound
axioms, we preserve the core version of the compiler, which does not
contain these axioms (or the problematic floating point operations),
verifying the unconditional correctness of our base compiler, as described in this paper.
Both versions of the compiler are provided in the artifact.

While IEEE floating point is fundamentally ill-suited for a bag
semantics, we hope that future work can explore some of the other
tradeoffs discussed above.

\subsection{Translation of the environment}
\label{sec:sqlatonrae:translation:envtran}

We have seen in~\Cref{sec:preliminaries:sqlalg} that, during the
evaluation of \sqlalg nested queries, one has to know groups and data of
outer queries: this is the role of the environment. This environment is
implicit: it is progressively populated when traversing queries, and the
correct slice in which to find the data (through attribute names) is
automatically computed when needed.

\nrae also has an ambient environment $\envnraenv$. However, its
manipulation is explicit: one has to store and retrieve data in it
through dedicated constructs.

To faithfully capture the \sqlalg semantics in \nrae, the implicit
manipulation of the \sqlalg environment thus has to be made explicit at
compile time. It means that the translation, in addition to reflecting
the query operators (see next section), adds administrative expressions
to manipulate the environment.

\paragraph{Runtime environment}

The runtime environment to execute a \nrae query coming from the
translation of a \sqlalg query mimics the \sqlalg environment. It has
the structure of a stack of slices, encoded as a link list in the record
$\envnraenv$:
$\envnraenv = \{\mathit{slice} : data_1, \mathit{tail} : \{\mathit{slice}: data_2, \mathit{tail} : ...\} \}$.
The values $data_1$, $data_2$, \dots correspond to each slice computed at runtime.

\paragraph{\nrae expressions for administrative steps}

For this environment to be correctly handled during the evaluation of
the query, the translation has to make explicit:
\begin{itemize}
\item how to populate $\envnraenv$ when traversing queries; and
\item how to retrieve data in the correct slice of $\envnraenv$.
\end{itemize}

For the populating part, translation will inject administrative steps to
add a new slice. We have seen in Section~\Cref{sec:preliminaries:sqlalg}
that slices can be composed of a single data or a collection. The
administrative steps are respectively these two \nrae expressions:
\[
  \begin{array}{lcl}
    \pushd
    & =
    & \{\mathit{slice}:\dbag{\qID},\mathit{tail}:\qENV\} \\
    \pushq
    & =
    & \{\mathit{slice}:\qID,\mathit{tail}:\qENV\} \\
  \end{array}
\]
We remind the reader that $\qENV$ is the expression that gives access to $\envnraenv$, and
$\qID$ is the expression corresponding to the current input data. Hence both constructions add the
current data on top of the current environment, with the difference that
for $\pushd$ the current data is put in a singleton collection.

To retrieve data in the nth slice of the environment, the administrative
step is simply the \nrae expression
$\qENV.\mathit{tail}.\cdots.\mathit{tail}.\mathit{slice}$ where there are
$n-1$ $\mathit{tail}$ projections.

\paragraph{Adding administrative steps at compile time}

To correctly add these steps at compile time, the translation function
for queries~$\Tau^{\mathsf{\algquery}}_{\aenv}$ is parameterized by a
\emph{abstract translation environment} $\aenv$, which contains static information
about the environment~$\env$. Indeed, in a slice~$(A,G,T)$, $A$ and $G$
depend only on the structure of the query. For example, for the query
${{\large \sigma}_f({\large \pi_{x \mathtt{\,as\,} a}(t)})}$, the top
slice~$S$ of the environment in which the formula $f$ is executed is
such that $A(S) = a$ and $G(S) = []$. Therefore, we define the
translation environment
$\aenv = [ S^{a}_n; ...; S^{a}_1]$ as a stack of static slices
$S^a_i = (A_i, G_i)$.

\begin{example}
Let us give the intuition on the \sqlalg queries
of~\Cref{sec:preliminaries:sqlalg:example}.\footnote{More details on this
  example are given at the end of the whole section.}

At compile time, the translation starts with an empty translation
environment.

It generates code for the outer ${\large\gamma}$ by
\begin{itemize}
\item inserting the \nrae code that populates the environment using
  the function $\pushq$; and
\item calling itself recursively on the translation environment
$[([a_1, b_1], a_1)]$ and the formula $\mathtt{exists}(...)$.
\end{itemize}
In this recursive call, similarly, it generates code by using $\pushq$
again and calling itself recursively on the translation environment
$[([a_2, b_2], a_2); ([a_1, b_1], a_1)]$ and the formula
$\mathtt{sum}(1+0*b) = 2$.

Finally, the translation of $b$ uses the translation environment to
insert the correct code to retrieve data: in the case of $b = b_2$, the code
is $\qENV.\mathit{slice}$ (since $b_2$ is in the first slice on the
translation environment); in the case of $b = b_1$, the code is
$\qENV.\mathit{tail}.\mathit{slice}$ (since $b_1$ is in the second
slice). \hfill $\blacksquare$
\end{example}

\paragraph{Proof invariant}

For the correctness statement and proof, we have to relate the \sqlalg
environment with the translation and runtime environments of \nrae. This
is done through two helper functions:
\begin{itemize}
\item the function $\taenv{\bullet}$ computes the translation
  environment by erasing the field~$T$ from \sqlalg environment's
  slices:
\[
\begin{array}{rcl}
\taenv{\tt []} & = & []\\
\taenv{(A,G,T)::\env} & = & (A,G)::\taenv{\env}
\end{array}
\]
\item the function $\tenv{\bullet}$ computes the runtime environment by erasing
  the fields~$A$ and $G$, and using the $\mathit{slice}$/$\mathit{tail}$
  records:
\[
\begin{array}{rcl}
\tenv{\tt []} & = & \{ \}\\
\tenv{(A,G,T)::\env} & = & \{\mathit{slice}: \trbag{T}, \mathit{tail} : \tenv{\env}\}
\end{array}
\]
\end{itemize}
These two functions are used only for specification and proofs, not
during the translation.

\subsection{Transformation of queries, formulas, and expressions}

\begin{figure*}[t]
  {\small
      \[
        \setstretch{1.3}
        \begin{array}{lcl}
          \trq{\aenv}{Q_1~~{\Large \bowtie}~~Q_2} & =
          & \trq{\aenv}{Q_1} \times \trq{\aenv}{Q_2}\\
          \trq{\aenv}{{\Large \sigma}_{f}(Q) } & =
          &  \qselect{\eapp{\pushd}{\trf{(\mathit{sort}~{Q},\{\})::\aenv}{f}}}
            {\trq{\aenv}{Q}}\\
          \trq{\aenv}{{\Large \pi}_{(\overline{e_n \mathtt{\,as\,} \attribute_n})} }(Q) & =
          & \qmap{\eapp{\pushd}
            {\trsel{(\mathit{sort}~{Q},\{\})::\aenv}{\overline{e_n \mathtt{\,as\,} \attribute_n}}}}
            {\trq{\aenv}{Q}} \\
           & & \\
          \trq{\aenv}{{\large \gamma}_{(\overline{e_n \mathtt{\,as\,} \attribute_n},\overline{b_k},f)}(Q) } & =
          & \text{let groups =}
            \qmap{\tdot{\qID}{g}}{\qgroupby{g}{\overline{b_k}}{\trq{\aenv}{Q}}}
            \text{in} \\
          & & \text{let filtered\_groups} = \qselect{\eapp{\pushq}
          {\trf{(\mathit{sort}~{Q},\overline{b_k})::\aenv}{f}}}
          {\text{groups}} \text{in} \\
          & &
          \qmap{\eapp{\pushq}
          {\trsel{(\mathit{sort}~{Q},\overline{b_k})::\aenv}{\overline{e_n \mathtt{\,as\,} \attribute_n}}}}
          {\text{filtered\_groups}}\\
          \multicolumn{3}{l}{\hspace*{9cm}\mbox{where}~g~
          \mbox{is a fresh label \wrt}~\overline{b_k}}\\
\end{array}
      \]
    }

   \caption{Non-trivial transformations of \sqlalg queries to \nrae.}\label{fig:qtrans}
\end{figure*}

\paragraph{Queries}\label{sec:sqlatonrae:translation:queries}
Translation of queries is denoted by $\trq{\aenv}{\_}$.
As explained above, the translation is parameterized by a translation environment~($\aenv$).
The translation of tables and set operations ($\mathtt{union}$, $\mathtt{intersect}$ and $\mathtt{except}$) is straightforward: they are simply
translated into the same operation in \nrae.
The translation of other queries is shown in \Cref{fig:qtrans}.

The join~($\bowtie$) is translated into a Cartesian product because \sqlalg queries have by construction distinct attribute names and in this case the semantics of both operators matches.

The translation of ${\Large \sigma}_{f}(Q)$ needs to take into account
the encoding of the \sqlalg environment in \nrae. It is translated as a
\nrae selection over the translation of $Q$, $\trq{\aenv}{Q}$. This
selection is performed over the translation of the formula~$f$ which, in
order to operate over the tuples of $\trq{\aenv}{Q}$, is computed in a
translation environment extended with the attributes introduced by
$Q$~(the {\em sort} of $Q$). An administrative step ($\pushd$) ensures
that the resulting \nrae query will be evaluated in a runtime
environment extended with the result of the evaluation of
$\trq{\aenv}{Q}$.

The translation of
${{\Large \pi}_{(\overline{e_n \mathtt{\,as\,} \attribute_n})}(Q)}$
is similar. The difference is that it is translated into the mapping
operator of \nrae. The pairs given to this operator are computed
recursively by translating each element: the function
$\Tau^{\mathsf{Sel}}_{\aenv}$ translates each expression $e_n$ and puts
them in a record.
$$
\trsel{\aenv}{\overline{e_n \mathtt{\,as\,} \attribute_n}} =
\{ \overline{\tratt{a_n}: \tre{\aenv}{e_n}} \}
$$

We now come to ${\large \gamma}$. We remind the reader that the query
${\large \gamma}_{(\overline{e_n \mathtt{\,as\,} \attribute_n},\overline{b_k},f)}(Q)$
performs three successive operations on $Q$: it first creates groups
using $\overline{b_k}$, then filters out some of these groups with
respect to the formula $f$, finally projects over expressions
$\overline{e_n}$ (giving names $\overline{\attribute_n}$). This order is
reflected in the translation. On the translation of $Q$,
$\trq{\aenv}{Q}$, it first creates groups using the helper function
$\ensuremath{\kkeyword{group\_by}}$, adding an extra column to remember
the grouping attribute thanks to the mapping
$\chi_{\left\langle{\tdot{\qID}{g}}\right\rangle}$.
Note that the grouping expressions must be attribute names, but it does
not reduce the expressiveness of \sqlalg. Second, the filtering is
performed, similarly as for the selection operator, except that it
applies to groups, meaning that the environment has groups instead of
singletons: the translation environment contains the groups
$\overline{b_k}$, and the runtime environment is extended using
$\pushq$. Finally, the projection is translated similarly to the
projection operator, except that it operates over groups as well.

\medskip

Some basic optimizations are also implemented during the translation. For
example, the selection over the $\mathtt{true}$ formula~(which is often
introduced by the pre-processing step from \sql to \sqlcoq) is directly simplified:
\[
      \begin{array}{l@{\,\,\,}c@{\,\,\,}l}
        \trq{\aenv}{{\Large \sigma}_{{\tt true}}(Q) }
        &=
        & {\trq{\aenv}{Q}}\\
        \trq{\aenv}{{\large \gamma}_{(\overline{e_n \mathtt{\,as\,} \attribute_n},\overline{b_k},{\tt true})}(Q) }
        & =
        &\qmap{\eapp{\pushq}
          {\trsel{(\mathit{sort}~{Q},\overline{b_k})::\aenv}{\overline{e_n \mathtt{\,as\,} \attribute_n}}}}
          {\qmap{\tdot{\qID}{g}}{\qgroupby{g}{\overline{b_k}}{\trq{\aenv}{Q}}}}
  \end{array}
      \]

\paragraph{Formulas}\label{sec:sqlatonrae:translation:form2n}

The main point in this translation is the use of three-valued logic, where Booleans are encoded with values of type $\typeeither$.
Each logical connective is thus translated into a \nrae expression
implementing the connective for three-valued logic~(\Cref{sec:sqlatonrae:datamodel}).

\paragraph{Expressions}\label{sec:sqlatonrae:translation:exptran}

\setlength{\tabcolsep}{2pt}
\begin{figure}
  
{\small
      \[
        \setstretch{1.3}
        \begin{array}{lcl}
          \trft{\aenv}{v} & = & \trval{v} \hfill \hspace*{0.6cm}
                                \mbox{if $v$ is a value}\\
\trft{(A, G) :: \aenv}{a}
                          & = &
                                \tdot{{\tt first\_elt\_of}
                                (\tdot{\qENV}
                                {\mathit{slice}})}{\tratt{a}} \hfill
                                \hspace*{0.2cm}
                                \mbox{if~} a \in A \\
          \trft{(A, G) :: \aenv}{a}
                          & = & \trft{\aenv}{a}~\circ^{e}
                                ~(\tdot{\qENV}{\mathit{tail}})
                                \hfill \hspace*{0.2cm}
                                \mbox{if~} a \notin A \\
          \trft{\aenv}{\fn~(\overline{\ef})}
                          & = & \trfn{\fn}
                                {(\overline{\trft{\aenv}{\ef}})} \\
          \tra{\aenv}{\fn~(\overline{\ea})}
                          & = & \trfn{\fn}
                                {(\overline{\tra{\aenv}{\ea}})} \\
          \tra{\aenv}{\ag(\ef)}
                          & = &
                                \eapp{({\tt remove\_slices}
                                ~(\aenv,\ef))}
                                {\trag{\ag}
                                {(\overline{\trf{((A,G) :: \aenv')}
                                {\ef}})}}\\
          \multicolumn{3}{l}{\qquad\mbox{where~}
            {\tt remove\_slices}~(\aenv,\ef)
            ~~\mbox{removes the same number of slices as~}
            \findevalaenv{\aenv}{\ef}}
        \end{array}
      \]
    }

   \caption{\sqlalg Expressions translation}\label{fig:etrans}
\end{figure}

\Cref{fig:etrans} defines the translation of expressions.
The translation of an attribute~$a$ reflects the use of the translation
environment that we have explained:
access to the correct slice in the runtime environment will be ensured by this translation.
If the attribute $a$ is in the top-slice~(\ie $a$ is in the set of labels defined in the slice), the value of the slice is extracted from the environment~($\tdot{\qENV}{\mathit{slice}}$).
Since elements in a slice are wrapped in a singleton collection~(\textit{c.f.}, $\pushd$), the function ${\tt first\_elt\_of}$ accesses the (only) element in the slice.
If the attribute $a$ is not in the top slice, the translated expression
removes the top slice~($\tdot{\qENV}{\mathit{tail}}$), allowing $a$ to
be accessed from the rest of the stack; accordingly, this slice is also
removed from the translation environment.

The translation of functions~($\fn$) and aggregates~($\ag$) requires handling \null values.
For some symbols, \null values are absorbing elements: if any input is
\null, the output is also \null. For example $q_1 + q_2$ is \null if
either $q_1$ or $q_2$ is \null.  Other symbols are neutral: they skip
\null values.  For example, \Sql{sum} $q$ will ignore any \null
elements in the bag returned by $q$.  Accounting for these different
behaviors correctly is not difficult, but needs to
be done carefully.
In addition, the translation of aggregates~($\ag$) has to access the
right slice in the environment. The number of slices to remove on top of
the environment stack is computed using the predicate
$\findevalaenv{\aenv}{\ef}$ which can be defined similarly to
$\findevalenv{\env}{\ef}$~(defined in \Cref{fig:sqlalg-semantics-summary}) since it does
not use the component $T$ of the slices of $\env$ (that is to say,
$\forall \env,\findevalaenv{\taenv{\env}}{\ef} =
\findevalenv{\env}{\ef}$).

\begin{example}
Let us illustrate the translation of the outer~${\large\gamma}$ in the
query \Sql{Q1} (or \Sql{Q2})
from~\Cref{sec:preliminaries:sqlalg:example}.
This translation first produces an \nrae
expression that builds the groups according to the grouping label using
$\kkeyword{group\_by}$ and discards the grouping keys to keep only the
groups:
$$
\text{groups} = \qmap{\tdot{\qID}{g}}{\qgroupby{g}{a_1}{t_1}}
$$
Once the groups are built, they are filtered using the formula~$\kkeyword{exists}(...)$.
Each group is put on the environment stack using~$\pushq$ for the execution of the formula.
The translation of the formula is recursively done in the environment
$\aenv = [([a_1, b_1], a_1)]$ reflecting the content of the stack, and
$\kkeyword{exists}(...)$ is translated into $\kkeyword{count}(...) >
0$:\footnote{We do not detail the translation of the condition of the $\kkeyword{exists}$ since it is very similar and would obscure the discourse.}
$$
\text{filtered\_groups} = \qselect{\eapp{\pushq}{(\kkeyword{count}(...) > 0)}}{\text{groups}}
$$
The last part of the translation is to project the parts of the groups that we are interested in.
The value of $\text{groups}$ is put on the top of the environment stack and
the translation of the projection is done in the same environment
$\aenv = [([a_1, b_1], a_1)]$. Since~$a_1$ is in the top slice of~$\aenv$, the
access of~$a_1$ becomes
$\tdot{\kkeyword{first\_elt\_of}(\tdot{\qENV}{\mathit{slice}})}{a_1}$.
The generated code is thus:
$$
\qmap{ \{ a_1 : \tdot{\kkeyword{first\_elt\_of}
                                (\tdot{\qENV}
                                {\mathit{slice}})}{x} \}}{\text{filtered\_groups}}
$$
$\blacksquare$
\end{example}

Putting it all together yields a fully certified compiler from \sqlalg
to \nrae, which handles most constructs of \sql (correlated queries,
\null values, most predicate, function and aggregate symbols). It enjoys
two variants: one without floating point values, and one with floating
point values that will be directly used in our target language,
JavaScript, but under invalid assumptions reflecting its incompatibility
with bag semantics.

\section{From \nrae to \imp}\label{sec:nnrctoimp}
\lstdefinelanguage{QCERTL}{
  basicstyle=\ttfamily,
}
\def\qcertl{\lstinline[basicstyle=\normalsize\ttfamily]}

Given the \NRAEnv intermediate language, we want to compile to it \js.
It requires (1)~a paradigm switch from functional to imperative, and (2)~a data representation switch from the internal data representation to JSON.

\begin{figure*}
  \centering
  \newcommand{\coqHTMLBase}{https://querycert.github.io}
\newcommand{\coqBaseModule}{html}
\newcommand{\coqurl}[3]{\coqHTMLBase/\coqBaseModule/Qcert.#1.#2.html\##3}

\tikzset{
    hyperlink node/.style={
        alias=sourcenode,
        append after command={
            let \p1 = (sourcenode.north west),
                \p2=(sourcenode.south east),
                \n1={\x2-\x1},
                \n2={\y1-\y2} in
            node [inner sep=0pt, outer sep=0pt,anchor=north west,at=(\p1)] {\href{#1}{\XeTeXLinkBox{\phantom{\rule{\n1}{\n2}}}}}
}
    }
}

\definecolor{cjava}{RGB}{156,186,95}
\definecolor{ccoqp}{RGB}{222,168,167}
\definecolor{ccoqc}{RGB}{161,187,215}

\tikzstyle{coqp}=[
 draw=ccoqp!150,fill=ccoqp
]

\tikzstyle{coqc}=[
 draw=ccoqc!150,fill=ccoqc
]

\tikzstyle{java}=[
 draw=cjava!150,fill=cjava
]

\tikzstyle{java-api}=[
 draw=cjava,fill=cjava!50,
 minimum width=21em,
 minimum height=1.4em,
 xshift=5em,
 font=\scriptsize\sffamily,
]

\tikzstyle{source}=[
 text width=2.75em,
 align=right
]

\tikzstyle{target}=[
 text width=4em,
 align=left
]

\tikzstyle{lang}=[
 thick,
 minimum width=3em,
 minimum height=1.4em,
 font=\scriptsize\sffamily,
]

\tikzstyle{sep-caption}=[
 thick,
 font=\scriptsize\sffamily,
]

\tikzstyle{tcoqp}=[
 draw=ccoqp!150
]

\tikzstyle{tcoqc}=[
 draw=ccoqc!150
]

\tikzstyle{trans}=[
 draw,semithick,-latex,
 font=\scriptsize\sffamily,
]

\tikzstyle{separation}=[
 draw,very thick, gray
]

\newcommand{\tikzoptim}[3][]{
   \path[trans, draw=#1!150, fill=#1!150] (#2) edge [looseness=10, loop above]
       node [pos=0.5](#2-optim) {}
       node [yshift = 0.5em,below] {~} ();
}
\newcommand{\tikzoptimdotted}[3][]{
   \path[trans, draw=#1!150, fill=#1!150] (#2) edge [looseness=10, loop above, dotted]
       node [pos=0.5](#2-optim) {}
       node [yshift = 0.5em,below] {{#3}} ();
}

\newcommand{\transref}[2]{
}

\newcommand{\transocamljava}[1]{
 \path[draw,thick,latex-latex, densely dotted,
      shorten >=0.2em, shorten <=0.2em,
       java] ([xshift=-0.75em] #1.south) -- ([xshift=-0.75em] #1-java.north);
}

 \begin{tikzpicture}[
   align=center,
   node distance=0.5em and 0.75em,
   font=\footnotesize,
   every loop/.style={latex-},
 ]

\node[lang, coqp, ]
      (nraenv) {\NRAEnv};

 \node[lang, coqp, right=of nraenv]
      (nnrc) {NNRC};

\node[lang, coqp, right=of nnrc]
      (nnrs) {NNRS};

\node[lang, coqp, right=of nnrs]
      (nnrsimp) {NNRSimp};

\node[lang, coqp, right=of nnrsimp]
      (impdata) {Imp(Data)};

\node[lang, coqp, right=of impdata]
      (impejson) {Imp(EJSON)};

\path[trans, tcoqp] (nraenv.east) -- (nnrc.west)
   \transref{3}{\coqurl{Translation}{NRAEnvtoNNRC}{nraenv_to_nnrc_top}};

 \tikzoptim[ccoqp]{nnrc}{NNRC(stratified)}
 \path[trans, tcoqp] (nnrc.east) -- (nnrs.west)
   \transref{3}{\coqurl{Translation}{NNRCtoNNRS}{nnrc_to_nnrs_top}};

\tikzoptim[ccoqp]{nnrs}{NNRS(no-shadow)}
 \path[trans, tcoqp] (nnrs.east) -- (nnrsimp.west)
   \transref{3}{\coqurl{Translation}{NNRStoNNRSimp}{nnrs_to_nnrs_imp_top}};

\path[trans, tcoqp] (nnrsimp.east) -- (impdata.west)
   \transref{3}{\coqurl{Translation}{NNRSimptoImpData}{nnrs_imp_to_imp_qcert_top}};

\path[trans, tcoqp] (impdata.east) -- (impejson.west)
   \transref{3}{\coqurl{Translation}{ImpDatatoImpEJson}{imp_qcert_to_imp_json}};

\end{tikzpicture}

   \caption{Compiler Pipeline}
  \label{fig:compiler-pipeline}
\vspace{-1.3em}
\end{figure*}
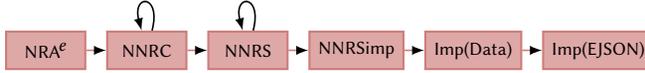

The correctness proof of this translation is the most challenging of the compilation chain.
In order to handle it, we decomposed the translation into the pipeline given in \Cref{fig:compiler-pipeline}.
It alternates source-to-source transformations and changes of intermediate languages where each step lowers some of the \NRAEnv constructs into simpler constructs that are closer to \js.
On the one hand, source-to-source transformations are simpler since they
allow us to deal with only one semantics at a time.
On the other hand, each intermediate language can enforce in its syntax and semantics some invariants which limit the scope of the proof.
The alternation of these two techniques allows us to make the transformation from a language to the next one simpler.

\subsection{From \nrae to NNRC}
\label{sec:nnrc}

Following the lead of \citet{DBLP:conf/sigmod/AuerbachHMSS17a}, we
first translate \NRAEnv to the named nested relation calculus (NNRC).
This calculus (an extension of~\citet{BusscheV07}), eliminates the implicit input
of \NRAEnv combinators, instead using explicit variables and
environments.  It also looks closer to a standard calculus for a
functional (bag-oriented) language.  As in the previous work, this
translation (and the accompanying translation between the languages'
associated type systems) are verified correct.

\subsection{From NNRC to NNRS}
\label{sec:nnrs}

NNRC, like \NRAEnv, is an expression oriented language: every
construct returns a value.  Many languages we would like to target, in
contrast, are statement oriented, and evaluation proceeds via
side-effects to variables. While \js supports expression oriented programming,
notably using first class functions, we would prefer a simpler
translation that, for example, can use a \qcertl{for} loop to express
iterators.
The next language, NNRS, is a statement oriented language: statements do not return values, but instead update the current state via (limited) side-effects.
This language is inspired by the normal form in the compilation of synchrous dataflow languages that identifes functional expressions that are translated into mutable ones~\citep{biernacki08}.

The translation from NNRC to NNRS is done in two steps.
First, we define NNRC(stratified), a subset of NNRC which distinguishes between basic expressions and complex expressions, and ensures that basic expressions never have complex sub-expressions.
Translation from NNRC to NNRC(stratified) hoists complex sub-expressions out of basic expressions by adding \qcertl{let} construct when needed.
For example, $\texttt{length}(\{(x+3) | x \in y\})$ is translated to
$\texttt{let } t_1 = \{(x+3) | x \in y\} \texttt{ in } \texttt{length}(t_1)$.
The definition of NNRC(stratified) in Coq is as a predicate on NNRC, so
the translation from NNRC to NNRC(stratified) is a source-to-source transformation.

The second step of the translation is the compilation of NNRC(stratified) to NNRS where complex expressions become statements.
NNRS introduces two forms of mutable variables: mutable data variables and mutable collection variables.
Both of them enforce a \emph{phase distinction}: in the first
phase, the mutable variable can (only) be updated, and in the second
phase, it can (only) be read.  In the first phase, mutable data
variables can be written (and re-written), and mutable collection
variables can have elements pushed (appended).  This phase distinction
is enforced using a form of \qcertl{let}, called \qcertl{LetMut} and \qcertl{LetMutColl}
respectively.  They each take a variable name and two statements.
They evaluate the first statement with the named variable being
mutable/appendable. The value of the variable is then frozen, and can
be read (but not mutated) by the second statement.  Reading from a frozen
data collection variable returns the accumulated bag.
This phase distinction avoids \emph{aliasing problems} by construction:
once we can read a variable, we can no longer modify it.

  In this language, \qcertl{for} loops no longer act as implicit maps:
  statements do not return values.  They are instead like \qcertl{for} loops
  (over a bag) in more traditional statement oriented/imperative
  languages.

  The translation from NNRC(stratified) to NNRS uses a form of
  continuation passing style, keeping track of a return continuation
  indicating which variable should store the return value. The
  \qcertl{let} statements are translated into mutable \qcertl{let} statements, where
  the final return, instead of being returned, is instead assigned to
  the variable. The \qcertl{for} loops are translated into a definition of a
  mutable collection variable, with the loop nested in the first
  branch of the mutable collection \qcertl{let} statement~(before the phase
  barrier).  The value returned by the body is pushed to the variable.

  If we continue the example of the compilation of
$\texttt{let } t_1 = \{(x+3) | x \in y\} \texttt{ in } \texttt{length}(t_1)$,
  the corresponding NNRS code (after some simplification) is:
\begin{jslisting}
letMutColl t1 from { for (x in y) { push(t1, x + 3); } };
return (length(t1))
\end{jslisting}
The \lstjs+letMutColl t1 from { ... }; ...+ constructs can update
\lstjs+t1+ in the block following the \lstjs+from+ and only
read its value after the \lstjs+return+.

\subsection{From NNRS to NNRSimp}
\label{sec:nnrsimp}

NNRS supports a limited form of side effects.
This suffices as a translation target for NNRC(stratified), but differs from target languages like \js.
In particular, it has three distinct namespaces, for different types of variables: mutable variables, mutable collections, and immutable variables.
Also, mutable \qcertl{let} statements put a variable in
different environments before and after the phase barrier (moving from the mutable data or collection namespace to the ``immutable'' namespace).

Separating these namespaces simplifies the translation from NNRC(stratified).
Notably, the different namespaces make it easy to pick fresh variables and ensure that no side effects are done on a variable after it is read.
But, the benefit of the three namespaces of NNRS also comes at a cost since we want to target languages with only one namespace.

The next language in the pipeline, NNRSimp, removes the features of NNRS that were introduced only to simplify the proofs, namely the phase distinctions and the separated namespaces of NNRS.
In NNRSimp, all variables are mutable (and readable).
There is a single \qcertl{let} construct (which introduces mutable variables), and a single assignment operator.

Similarly to the compilation from NNRC to NNRS, we first define a subset of the source language, NNRS(no-shadow), to simplify the compilation to the target language, NNRSimp.
We define a predicate, named \emph{cross-shadow-free}, that specifies what name conflicts are problematic.
The intuition is that traditional shadowing is still ok, but shadowing across namespaces causes problems when they are conflated.
We define a source-to-source transformation that renames variables to ensure that the result is cross-shadow-free.
The translation is idempotent, and tries to rename variables minimally.
Of course, it is verified to be semantics and type preserving.

Once a program is in NNRS(no-shadow) form, it is compiled to NNRSimp.
The NNRS(no-shadow) language ensures that no false shadowing conflicts are introduced when the three namespaces of NNRS are collapsed into one namespace.
Immutable \qcertl{let} statements are re-written to be mutable \qcertl{let} statements, which happen to mutate the value at most once.
Mutable collection variables are encoded by initializing a mutable variable with the empty bag.

\subsection{From NNRSimp to Imp}
\label{sec:impejson}

The final language in the compilation chain is \imp, which is used to handle the switch in data representation.
As presented in \Cref{sec:preliminaries:imp}, the \imp language is parameterized
by its data model and the operations~on it. We take advantage of that
by compiling NNRSimp to Imp in two steps: first we translate to
Imp(Data), which preserves the \nrae{} data
model~(\Cref{sec:preliminaries:nrae}). We then translate Imp(Data) to Imp(EJson), which still uses Imp,
but over a (slightly extended) JSON data~model.

\paragraph*{NNRSimp to Imp(Data)}

The main difference between NNRSimp and Imp(Data) is the lack of an
Imp language construct for pattern matching on values of type
\typeeither{}. This NNRSimp construct is compiled into an
\lstinline+if/then/else+ using an Imp function \qcertl{either} to test
if a value is a \qcertl{left} or not, and then using \qcertl{getLeft}
and \qcertl{getRight} functions to deconstruct \typeeither{} values
appropriately.

The operators of Imp(Data) are the same as the previous languages and
the library functions are \qcertl{either}, \qcertl{getLeft}, and
\qcertl{getRight}.
Finally, if the \NRAEnv \qcertl{group_by} construct was preserved (and not removed as described in \cref{sec:preliminaries:nrae}),
the library of Imp(Data) must provide a \qcertl{group_by} function.

\paragraph*{From Imp(Data) to Imp(EJson)}

Now that the query is in \imp, the last step is to switch data
models, from the one of \nrae{} to JSON. We use a small extension to
the official JSON representation by adding a biginteger type in
addition to \js numbers. This is necessary to preserve the semantics
for integer operations in \sql which in our formalization relies on
the \texttt{Z} Coq type.

The change of data model is fundamental in that it really introduces a
representation specific to the target language for the compiler (here
\js). In essence: collections are translated into \js arrays, records
are translated into \js objects, and the $\texttt{left}\ d$ and
$\texttt{right}\ d$ values of Data are encoded as JSON objects with
reserved names \lstjs+{ "$left":+~$d$~\lstinline+}+ and \lstjs+{ "$right":+~$d$~\lstinline+}+.

Imp(Data) functions on \qcertl{left} or
\qcertl{right} values must be translated into equivalent Imp(EJson)
functions on those objects, relying on JavaScript's ability to
check if an object has a specific property.

\paragraph*{Correctness}

The shape of the correctness theorem for the translation from
Imp(Data) to Imp(EJson) is worth mentioning. First it relies on a
translation function from Data to EJson with good properties. Notably
two Data values which translate to the same EJson have to be equal.

\begin{lstlisting}[language=SSR]
Lemma data_to_ejson_inj d1 d2: data_to_ejson d1 = data_to_ejson d2 -> d1 = d2.
\end{lstlisting}

This property is fundamental to proving the main correctness theorem:
\begin{lstlisting}[language=SSR]
Lemma imp_data_function_to_imp_ejson_function_aux_correct h (d:data) (f:imp_data_function) :
  lift data_to_ejson (imp_data_function_eval h f d) =
  imp_ejson_function_eval h (imp_data_function_to_imp_ejson f) (data_to_ejson d).
\end{lstlisting}
which states that evaluating an Imp(Data) function on some data
\texttt{d} and translating the result to EJson yields the same result
as evaluating the corresponding Imp(EJson) function on the translation
of \texttt{d} to EJson. Note that this formulation means the
correctness theorem only holds for evaluating Imp(EJson) values
\emph{resulting from translating a Data value}, not for arbitrary
EJson data. We believe this formulation provides the right invariant
for the compiler, but this imposes that at runtime only EJson values
that correspond to valid Data values are passed, a property we are
careful to ensure.  This additional constraint is needed (unlike our
earlier translations, which do not have such a constraint), because
the target data model is larger, and allows for invalid data.

\section{Implementation}\label{sec:implementation}
\dbcert is built from a certified core in Coq with additional
non-certified components in OCaml and JavaScript. We review those
components here.
The full development can be found in the artifact,
including some examples focused on testing the more subtle aspects of SQL's semantics~\cite{artifact}.

\subsection{The main theorem}
\label{sec:implementation:theorem}

The certified core links the various translations between intermediate
languages described in the previous sections. A theorem of semantics
preservation for the full pipeline is obtained by combining individual
translation proofs.

\begin{theorem}[Semantics preservation]
  Given a schema and a \sqlcoq query $Q$:
  \begin{itemize}
  \item {\bf if} $Q$ is well-formed, in the sense of
    \citet{DBLP:conf/cpp/BenzakenC19},
  \item {\bf then} the compiler outputs an Imp query $q$ such that, on every
    valid instance of the schema $i$, $\semq{Q}{}{i}$ is equal to
    $\impsem{q}{\tri{i}}$ (upto bag equality).
  \end{itemize}
\end{theorem}

\subsection{SQL parser}
\label{sec:implementation:parser}

The \dbcert implementation includes a SQL parser, written in OCaml,
which is used to construct an initial \sqlcoq abstract syntax tree
(AST). The SQL grammar is written using the menhir parser generator,
which can be used to generate the parser either using standard menhir
or its Coq back-end~\cite{DBLP:conf/esop/JourdanPL12} and extracting
the (thus proven complete) parser.

The initial construction of the \sqlcoq AST performs some simple
normalization of the SQL query ({\em e.g.} adding a \Sql{where true}
if the \Sql{where} clause is missing). It ensures every intermediate
expression has been named, yielding
well-formed \sqlcoq queries. It also ensures that all attribute
names are different, tagging them with the name of the relation
they belong to.  This step is not yet certified.

\subsection{JavaScript Code Generation}
\label{sec:implementation:js}

From the generated Imp(EJson) code, \dbcert creates a \js string in
two steps. First, Imp(EJson) is translated into a \js AST based on the
JSCert~\cite{DBLP:conf/popl/BodinCFGMNSS14} formalization of
JavaScript. Then, the JSCert AST is pretty-printed as a \js string.

The current \dbcert produces ECMAScript 6 compliant code. It uses
JavaScript blocks with \texttt{let} bindings to ensure that variable
scoping in Imp blocks is being preserved in the generated code. It
relies only on a small subset of ECMAScript 6 and should run in most
versions of Node.js and modern browsers. The artifact has
been tested with Node.js version 10.

\subsection{JavaScript Runtime}\label{sec:implementation:runtime}

Execution of SQL queries compiled to JavaScript with \dbcert relies on
a small run-time library written in JavaScript as well. This runtime
serves two purposes: it implements runtime functions specified by the
instantiation of \imp on EJson; it is used as a pre-processor for
the query input in JSON, and as a post-processor for the query output.

\paragraph{EJson runtime} EJson supports JavaScript numbers~(IEEE754
floating point numbers) and persistent (functional) arrays. The
runtime provides functions to manipulate these values.

For persistent arrays, we provide two implementations. Our initial
implementation was using \js arrays directly, with each array operation
creating a new array. But the most common operation used in the
compiled code is \lstjs+push+ which adds one element to an
array. Doing a copy of the entire array for every \lstjs+push+ has a strong
impact on performances.

To address that issue, our current implementation uses persistent
arrays where several arrays can be represented as views on the same
backing data. The goal is to keep the implementation of the runtime
simple and improve the performance of the \lstjs+push+ operation. A
persistent array is simply an object with two fields:
\lstjs+$content+, the \js array containing the data, and
\lstjs+$length+, an integer indicating the view of the array. With
this representation, the \lstjs+push+ operation can be implemented
such that adding an element requires a copy of the
data~(\lstjs+slice+) only if the size of the backing array differs
from the one stored in \lstjs+$length+.

\paragraph{Pre- and post-processors}
The runtime takes care of encoding JSON values into the expected
format for Imp(EJson). It includes the encoding of values which may or
may not be \null into the appropriate {\dleft{}} and {\dright{}}
representation used internally by \dbcert, as described in
\Cref{sec:sqlatonrae:datamodel} and \Cref{sec:impejson}.
It also renames record fields to be consistent with the renaming
applied when normalizing the \sqlcoq AST.
For instance, the following JSON input:
\begin{lstlisting}[language=SQLL]
{ "persons" : [ { "name" : "John" },  { "name" : null } ] }
\end{lstlisting}
is pre-processed to the following:
\begin{lstlisting}[language=SQLL]
{ "persons" : array( { "persons.name" : { "left" : "John" } },
                     { "persons.name" : { "right" : null } } ) }
\end{lstlisting}
where \lstinline{array} is the constructor for EJson persistent arrays.

\subsection{\dbcert Runner}
\label{sec:implementation:runner}

For convenience, we provide a small Node.js script which allows one to
execute queries compiled with \dbcert on JSON data. This script
performs the following tasks:
\begin{itemize*}
\item load a SQL query compiled to JavaScript through \dbcert;
\item load and pre-process the database in JSON format;
\item execute the query;
\item post-process and print the query result in JSON format.
\end{itemize*}

For instance, here is a (re-flowed) trace for the compilation and execution of an SQL query.
\begin{lstlisting}[language=SQLL, mathescape]
bash-3.2> cat tests/org2.sql 
create table employees (name text, age int);
select name from employees where age > 32;
bash-3.2> ./dbcert -link tests/org2.sql
Corresponding JS query generated in: tests/org2.js
Compilation to JavaScript finished
bash-3.2> cat tests/db1.json 
{ "employees": [ { "name" : "John", "age" : 34 }, { "name" : "Joan", "age" : 32 },
                 { "name" : "Jim", "age" : 33 }, { "name" : null, "age" : 35 },
                 { "name" : "Jill", "age" : null } ] }
bash-3.2> node ./dbcertRun.js tests/org2.js tests/db1.json
[{"name":"John"},{"name":"Jim"},{"name":null}]
\end{lstlisting}

\section{Evaluation and related work}\label{sec:related}
\subsection{Evaluation}
\label{sec:related:evaluation}
We compare \dbcert with AlaSQL~\cite{alasql},
Q*cert~\cite{DBLP:conf/sigmod/AuerbachHMSS17a}, and SQL.js~\cite{sqljs} which all execute \sql queries on \js.
\alasql is a popular \js library with more than $13$k weekly downloads
on \url{https://www.npmjs.com} and $5.6$k stars on GitHub.
The SQL compiler from Q*cert also produces \js. \dbcert uses the
translation from \nrae to \nnrc of this compiler. But compared to
\dbcert, Q*cert directly translates \sql to \nrae and produces \js code
directly from \nnrc. Both of these translation are not formally
verified, and in particular the translation from \sql to \nrae does not
correctly handle environments. \null values are not supported by this
compiler.
SQL.js is SQLite compiled to WebAssembly that can be then executed by the JavaScript engine.
It is thus directly based on the implementation SQLite, one of the most widely deployed implementation of SQL.

We evaluate the correctness of the compiler using the queries proposed by the papers of \citet{DBLP:journals/pvldb/GuagliardoL17}
and \citet{DBLP:conf/cpp/BenzakenC19}.
These queries have been designed to notably exercise the use of \null and correlated queries.
The difficulty with \null is that it is generally considered as different from every values, including itself, although it is sometimes considered equal to itself.
The challenge wih correlated queries is that the behavior of a subquery can depend on its evaluation context.

The benchmark contains a total fifteen queries. Four queries are covering \null values: three proposed by \citet{DBLP:journals/pvldb/GuagliardoL17} and one by \citet{DBLP:conf/cpp/BenzakenC19}. The remaining eleven queries are covering correlated queries and are proposed by \citet{DBLP:conf/cpp/BenzakenC19}.

We take as reference the answers given by the SQL standard (when precise enough), three major RDBMSs~(Oracle, PostgreSQL, SQLite), and
the formal semantics of~\citet{DBLP:conf/cpp/BenzakenC19}.
On the considered queries, all of these systems agree on the expected results.

All the queries and database instances used for the evaluation are provided in \ifextended \Cref{sec:evaluation-app} and in the artifact~\cite{artifact}\fi{}. We refer the readers to the original papers for additional details.

The following table summarizes the results: for each compiler, we
give number of valid answers per number of queries.\footnote{The incorrect behaviors have been reported in issues $1414$ and $1416$ on~\url{https://github.com/agershun/alasql/}.}

\begin{center}
\begin{tabular}{r@{\ \ \ }|@{\ \ \ }c@{\ \ \ }|@{\ \ \ }c@{\ \ \ }|@{\ \ \ }c@{\ \ \ }|@{\ \ \ }c}
    Benchmarks & \dbcert & AlaSQL & Q*cert & SQL.js \\
    \hline
    \null & 4/4 & 3/4 & N/A & 4/4 \\
    correlated queries & 11/11 & 7/11 & 9/11 & 11/11 \\
  \end{tabular}
\end{center}

We note that many SQL query compilers handle these
kinds of queries differently from the standard and well-established
RDBMSs. These differences may lead to subtle bugs, resulting in
corruption of data and processes.  It is crucial to ensure the
semantic correctness of compiled queries.

A preliminary performance evaluation of the generated code is presented
in \ifextended \Cref{sec:related:evaluation:performance}\fi{}, but a proper evaluation is left as future work.

\subsection{Challenges and methodology}
\label{sec:related:evaluation:challenges}

\paragraph{Translation from \sqlalg to \nrae}

\dbcert is built on top of two existing projects~\cite{DBLP:conf/cpp/BenzakenC19,DBLP:conf/sigmod/AuerbachHMSS17a} which made different design choices.
For example, they used different techniques to implement extensible data models.
In \sqlalg, the formalization has two levels: a generic specification level, and a realization level that instantiates the generic components with computational definitions.
In \nrae, the data model is concrete, with extension points for external data and operators abstracted using type classes.

Both approaches have some benefits and drawbacks.
To start with a project, the concrete approach of \nrae is easier, it provides concrete objects to think, execute, and debug, whereas the abstract approach of \sqlalg makes the concept more difficult to grasp and requires a realization of the data model to be able to experiment.
On the other hand, the abstract approach provides a nice uniform interface to select the data model.
The concrete approach necessitates splitting the code of some functions between the core of the data model and the instantiation of the extension point.

The different approaches employed created some challenges when connecting the two projects.
To preserve the genericity of \sqlalg with respect to the data model, the equivalence between data-models is first specified and then realized according to the concrete data model.
This separation introduced by the abstract approach helped the proof development by dividing it into two phases.

Both approaches successfully enabled adding float values to the data
model.  The validity of the core translation needed no modification.
When realizing the abstract model, the formalization revealed \sql's
weakness when specifying the summing and averaging of float values, as
discussed in Section~\ref{sec:sqlatonrae:datamodel:float}.

Regarding proofs, the most challenging one in this part was the
correctness of the translation of the environment: in \sqlalg, the whole
environment is present at runtime; whereas in \nrae, the static part of
the environment has been embedded in the query during the translation,
and only the dynamic part is present at runtime. Relating the two in the
induction was demanding.
Another exigent proof was the correctness of the translation of \null
values and three-valued logic. We established that the chosen \nrae
operators correctly implemented the \sqlalg operators. Coq was a
particularly effective tool in this context, as the translation itself
could be guided by the proof.

\paragraph{\dbcert back-end}

Building the \dbcert back-end involved solving three major and quite
different hurdles: the paradigm switch from functional to imperative
languages, switching the data representation from relations to JSON,
and handling variable names and scoping. We used a few specific
strategies in order to deal with these difficulties while enabling
proof development.

First, we used a large number of intermediate languages. This allows us to
enforce some invariants in the syntax and semantics of each language,
tackling each hurdle one at a time. We are satisfied by this
approach, which simplifies the proofs and limits their scope. The
additional translation phases do not seem to negatively affect the
extracted compiler, with most of the compilation time spent on
optimization.
The proliferation of intermediate languages does have the
disadvantage of increasing the size of the code base. But the presence
of proofs simplifies maintenance since any breaking change in the
code is immediately detected when compiling the corresponding proofs.

Second, we used small languages to keep them simple. This choice
sometimes leads to complex encodings of some language constructs
into the next one.
As an example, \nrae does not have an \lstjs+if+ construct. This has
little impact on how we write and prove the translation: we write in
Coq a function that generates a conditional using a
selection~($\sigma$), prove that it behaves like a \lstjs+if+, and
then use it instead of a \nrae language construct.  However, this does
impact the generated code, introducing redundant packing and unpacking
of data in collections.  These then need to be simplified through
optimizations.
Using small languages can thus result in a more complex compiler,
despite simplifying the functions and properties of each individual
language.

\subsection{Related work}
\label{sec:related:evaluation:related}

The very first attempt to verify a RDBMS, using \coq, is presented
in~\citet{popl10morisset}. The \sql fragment considered is a
reconstruction of \sql in which attributes are denoted by
position. Several key SQL features, such as \Sql{group by having}
clauses, quantifiers in formulas, nested, correlated queries,
\null's, and aggregates, are not covered. A tool to decide \sql query
equivalence was presented in~\citet{Chu:2017:HPQ:3062341.3062348}. It
relies on a K-relation~\cite{DBLP:conf/pods/GreenKT07} based semantics
for \sql which handles the \Sql{select from where} fragment with
aggregates but does not include \Sql{having} or handle \null
values. Like~\citet{popl10morisset}, they used a reconstruction of the
language, avoiding the trickier aspects of variable
binding. Additionally, their semantics are not executable, making it difficult
to compare it to other \sql implementations.

More closely related to our work, a translation from \sql to \nrae was
developed as part of a certified query compiler effort
in~\citet{DBLP:conf/sigmod/AuerbachHMSS17a}. That translation supports
a realistic subset of \sql, including notably \Sql{group by having},
but it did not handle null values. It also did not include a semantics
for \sql and the translation was therefore not proved correct. To the
best of our knowledge the most complete formal and mechanized
semantics for \sql is that developed
in~\citet{DBLP:conf/cpp/BenzakenC19}, notably covering most subtleties
of \sql for a practical fragment with nested correlated queries and null
values. While it is executable, which means it can be checked against
actual \sql implementations, it relies on a simple interpreter with no
compilation or algebraic optimization. Our work relies heavily on both
projects.

Since \dbcert compiles to JavaScript, it is relevant to discuss the
mechanized \js specification presented
in~\citet{DBLP:conf/popl/BodinCFGMNSS14}.
While our work uses the AST provided by their work~\cite{DBLP:conf/popl/BodinCFGMNSS14}
for the final code generation, attempting to prove that final part of the
translation correct with respect to their semantics is left as future work.

\section{Conclusion}\label{sec:concl}
We have presented a formally verified compiler from \sql to a general
purpose imperative language, with a \js back-end. \dbcert handles a
large subset of the \sql language, including nested queries and null
values. Most of the compiler was proved correct using the Coq
interactive theorem prover. The extracted compiler is fully functional
and produces portable \js code which can be executed in various
environments. Importantly, one of the intermediate representations is a
classic database algebra for which a large numbers of optimization
techniques have been
developed~\cite{DBLP:conf/dbpl/CluetM93,claussen1997optimizing,querycompilers}. We
believe this is an important step toward the development of a fully
certified and practical query compiler.

\bibliography{biblio}

\ifextended

\appendix
\newpage
\section{Presentation of the artifact}
\label{sec:supplementary}

The artifact~\cite{artifact} is a self-contained version of the implementation presented in this
article (including
work we build on). It is developed using Coq-8.11, OCaml and JavaScript.

A {\sf README.md} file indicates how to compile the code and experiment
with it. The code is divided into 6 directories:
\begin{itemize}
\item The directory {\sf datacert} contains the mechanized SQL semantics
  from~\citet{DBLP:conf/cpp/BenzakenC19}, augmented with floating point
  values and a SQL parser.
\item The directory {\sf qcert} contains the mechanized nested relation
  algebra from~\citet{DBLP:conf/sigmod/AuerbachHMSS17}, augmented with
  the new imperative backend presented in Section~\ref{sec:nnrctoimp}.
\item The directory {\sf jsql} contains the new certified compiler from
  SQLAlg to NRAe presented in Section~\ref{sec:sqlatonrae}.
\item The directory {\sf extraction} contains the new, simple OCaml glue
  to execute the extracted compiler presented in
  Sections~\ref{sec:implementation:parser} and
  \ref{sec:implementation:js}.
\item The directory {\sf runtime} contains the new JavaScript runtime
  used by the execution of the compiler, presented in
  Section~\ref{sec:implementation:runtime}.
\item The directory {\sf tests} contains examples and the experiments
  presented in Section~\ref{sec:related:evaluation}.
\end{itemize}

The important aspects of the development are presented in
Table~\ref{tab:supplementary}.

\begin{table}[!h]
\begin{tabular}{|l|l|l|l|}
\hline
{\bf Description} & {\bf Code} & {\bf Paper} & {\bf New}\\
\hline
Definition of SQLCoq & {\sf datacert/data/sql/Sql.v}, l.115 & & \\
Definition of SQLAlg  & {\sf datacert/data/sql/SqlAlgebra.v}, l.56 & \ref{sec:preliminaries:sqlalg} & \\
Definition of NRAe & {\sf qcert/compiler/core/NRAEnv/Lang/NRAenv.v}, l.66 & \ref{sec:preliminaries:nrae} & \\
Definition of Imp & {\sf qcert/compiler/core/Imp/Lang/Imp.v}, l.67 & \ref{sec:preliminaries:imp} & \checkmark \\
\hline
Translation from SQLAlg to NRAe: & {\sf jsql/} & \ref{sec:sqlatonrae} & \checkmark \\
- Correctness theorem & \hfill {\sf query/QueryToNRAEnv.v}, l.3115  & \ref{sec:sqlatonrae:correctness} & \checkmark \\
- Decoding theorem & \hfill {\sf decode/Decode.v}, l.88 & \ref{sec:sqlatonrae:correctness} & \checkmark \\
- Specification of the data model & \hfill {\sf data/TupleToData.v} & \ref{sec:sqlatonrae:datamodel} & \checkmark \\
- Realisation & \hfill {\sf query/TnullQN.v}, {\sf formula/TnullFN.v}, \dots & \ref{sec:sqlatonrae:datamodel} & \checkmark \\
- Axioms on floats & \hfill {\sf aux/AxiomFloat.v} & \ref{sec:sqlatonrae:datamodel:float} & \checkmark \\
- Translation of the instance & \hfill {\sf instance/InstanceToNRAEnv.v} & \ref{sec:sqlatonrae:datamodel} & \checkmark \\
- Translation of environments & \hfill {\sf env/EnvToNRAEnv.v} & \ref{sec:sqlatonrae:translation:envtran} & \checkmark \\
- Translation of queries & \hfill {\sf query/QueryToNRAEnv.v} & \ref{sec:sqlatonrae:translation:queries} & \checkmark \\
- Translation of formulas & \hfill {\sf formula/FormulaToNRAEnv.v} & \ref{sec:sqlatonrae:translation:form2n} & \checkmark \\
- Translation of expressions & \hfill {\sf term/FTermToNRAEnv.v}, {\sf term/ATermToNRAEnv.v} & \ref{sec:sqlatonrae:translation:exptran} & \checkmark \\
\hline
Translation from NRAe to Imp: & qcert/compiler/core/Translation/Lang/ & \ref{sec:nnrctoimp} & \checkmark \\
- NRAe to NNRC & \hfill {\sf NRAEnvtoNNRC.v} l.617 & \ref{sec:nnrc} & \\
- NNRC to NNRS & \hfill {\sf NNRCtoNNRS.v}, l.1063 & \ref{sec:nnrs} & \checkmark \\
- NNRS to NNRSimp & \hfill {\sf NNRStoNNRSimp.v}, l.692 & \ref{sec:nnrsimp} & \checkmark \\
- NNRSimp to Imp(Data) & \hfill {\sf NNRSimptoImpData.v}, l.487 & \ref{sec:impejson} & \checkmark \\
- Imp(Data) to Imp(EJSON) & \hfill {\sf ImpDatatoImpEJson.v}, l.1474 & \ref{sec:impejson} & \checkmark \\
\hline
Main theorem & {\sf jsql/poc/ToEJson.v}, l.175 & \ref{sec:implementation:theorem} & \checkmark \\
\hline
Runtime and glue: & & \ref{sec:implementation} & \checkmark \\
- SQL parser & {\sf datacert/plugins} & \ref{sec:implementation:parser} & \checkmark \\
- JavaScript backend & {\sf qcert/compiler/core/Translation/Lang/} &  &  \\
 & \hfill {\sf ImpEJsontoJavaScriptAst.v} & \ref{sec:implementation:js} & \checkmark \\
- JavaScript runtime & {\sf qcert/runtimes/javascript/qcert\_runtime.ml} & \ref{sec:implementation:runtime} & \checkmark \\
- DBCert runner & {\sf dbcertRun.js} & \ref{sec:implementation:runner} & \checkmark \\
\hline
Semantics benchmarks: & & \ref{sec:related:evaluation} & \\
- Examples with null values & {\sf tests/null} & \ref{subsec:examples}, \ref{sec:related:evaluation} & \\
- Examples with correlated queries & {\sf tests/nested} & \ref{subsec:examples}, \ref{sec:related:evaluation} & \\
Performance benchmarks: & & \ref{sec:related:evaluation:performance} & \\
- Database of 58,800 entries & {\sf tests/simple/orgX.sql}, {\sf tests/simple/db1big.json} & \ref{sec:related:evaluation:performance} & \\
\hline

\end{tabular}
\caption{Main references to the code development}
\label{tab:supplementary}
\end{table}

\newpage

\section{Semantics}
\label{sec:semantics-app}
This section provides the complete semantics of \sqlalg and \nrae.

\subsection{\sqlalg}
\label{sec:semantics-app:sqlalg}
The semantics of \sqlalg is defined in \Cref{fig:sqlalg-semantics,fig:sqlalg-formula-semantics,fig:algsem} and follows the one presented in~\citet{DBLP:conf/cpp/BenzakenC19}.
The semantics $\semq{Q}{\env}{i}$ of a query~$Q$ evaluated in an environment~$\env$ on a database instance $i$ defines a bag.
The instance~$i$ associates the data to each table.
The environment~$\env$ defines the local evaluation context of the query.

We refer the reader to~\citet{DBLP:conf/cpp/BenzakenC19} for a detailed explanation of the semantics.

\begin{figure}[!h]
  {\footnotesize
\[
\setstretch{1.3}
\begin{array}[t]{l@{~}c@{~}l}
\semq{\mbox{\tt ()}}{\env}{i} & = & \{\!| ~ |\!\} \\
\semq{\mathit{tbl}}{\env}{i}  & = & i.\mathit{tbl} \quad \mbox{~if~} \mathit{tbl} \mbox{~is a table}\\
\semq{Q_1\mathtt{~union~} Q_2}{\env}{i} & = & \semq{Q_1}{\env}{i}\cup\semq{Q_2}{\env}{i}\\
\semq{Q_1\mathtt{~intersect~} Q_2}{\env}{i} & = & \semq{Q_1}{\env}{i}\cap\semq{Q_2}{\env}{i}\\
\semq{Q_1\mathtt{~except~} Q_2}{\env}{i} & = & \semq{Q_1}{\env}{i}\setminus\semq{Q_2}{\env}{i}\\
\semq{Q_1 ~{\Large \bowtie}~ Q_2}{\env}{i} & = & \\
\multicolumn{3}{l}{\left\{\!\left|\left(\overline{a_n=c_n},\overline{b_k=d_k}\right)
\left|
\begin{array}{l}
(\overline{a_n=c_n}) \in \semq{Q_1}{\env}{i} ~\wedge\\
(\overline{b_k=d_k}) \in \semq{Q_2}{\env}{i}  ~\wedge\\
(\forall~ n,k, ~~ a_n = b_k \Rightarrow c_n = d_k)
\end{array}
\right.\right|\!\right\}} \\
\end{array}
\hspace*{-0.9em}
\begin{array}[t]{l@{~}c@{~}l}
\semq{{\large \pi}_{(\overline{e_n \mathtt{\,as\,} a_n})}(Q)}{\env}{i} & = &
\{\!| (\overline{a_n = \seme{e_n}{(\ell(t),[],[t]) :: \env}}) \mid
t \in \semq{Q}{\env}{i}
|\!\}\\
\semq{{\large \sigma}_{f}(Q)}{\env}{i} & = &
\{\!|t\in\semq{Q}{\env}{i} \mid \semf{f}{(\ell(t),[],[t]) :: \env}{i} = \top|\!\}\\
\multicolumn{3}{l}{
\semq{{\large \gamma}_{(\overline{e_k \mathtt{\,as\,} a_k},\overline{e_n},f) }(Q)}{\env}{i} = } \\
\multicolumn{3}{c}{\left\{\!\left| \overline{(a_k=\seme{e_k}{(\ell(T),\overline{e_n},T)::\env})} | T\in\mathbb{F}_3 \right|\!\right\}} \\
\multicolumn{3}{@{\hspace{1cm}}l}{\mbox{and~} \mathbb{F}_2 \mbox{~is a partition of $\semq{Q}{\env}{i}$ according to $\overline{e_n}$} }\\
\multicolumn{3}{@{\hspace{1cm}}l}{\mbox{and~} \mathbb{F}_3 = \left\{\!\left|T \in {\mathbb F}_2  \left| \semf{f}{(\ell(T),\overline{e_n},T)::\env}{i} = \top\right.\right|\!\right\}} \\
\end{array}
\]
}

 \caption{Semantics of \sqlalg queries.}
\label{fig:sqlalg-semantics}
\end{figure}

\begin{figure}[!h]
  {\footnotesize
\[
\setstretch{1.3}
\begin{array}[t]{l@{~}c@{~}l}
\semf{f_1 \mathtt{~and~} f_2}{\env}{i} & = & \semf{f_1}{\env}{i} \wedge_3 \semf{f_2}{\env}{i}\\
\semf{f_1 \mathtt{~or~} f_2}{\env}{i} & = & \semf{f_1}{\env}{i} \vee_3 \semf{f_2}{\env}{i}\\
\semf{\mathtt{not~} f}{\env}{i} & = & \neg{}_3 \semf{f}{\env}{i}\\
\semf{\mathtt{true}}{\env}{i} & = & \mathit{true}\\
\semf{p(\overline{e_n})}{\env}{i} & = & p(\overline{\sema{e_n}{\env}})\\
\end{array}
\quad
\begin{array}[t]{l@{~}c@{~}l}
\semf{p(\overline{e_n}, \mathtt{~all~} q)}{\env}{i} & = & \mathit{true}
~~{\mbox{iff~} \semf{p(\overline{e_n},t)}{\env}{i} = \mathit{true} \mbox{~for all~} t \in \semq{q}{\env}{i}}\\
\semf{p(\overline{e_n}, \mathtt{~any~} q)}{\env}{i} & = & \mathit{true}
~~{\mbox{iff~} \semf{p(\overline{e_n},t)}{\env}{i} = \mathit{true} \mbox{~and exists~} t \in \semq{q}{\env}{i}}\\
\semf{\overline{e_n \mathtt{~as~} a_n} \mathtt{~in~} q}{\env}{i} & = & \mathit{true}
~~{\mbox{if~} (\overline{a_n=\sema{e_n}{\env}}) \mbox{~belongs to~} \semq{q}{\env}{i}}\\
\semf{\mathtt{exists~} q}{\env}{i} & = & \mathit{true}
~~\mbox{iff~} \semq{q}{\env}{i} \mbox{~is not empty}
\end{array}
\]
}

 \caption{Semantics of \sqlalg formulas.}
\label{fig:sqlalg-formula-semantics}
\end{figure}

\begin{figure*}[!h]

\begin{minipage}[c]{.4\linewidth}
{\footnotesize
\[
\def\arraystretch{1.2}
\begin{array}{lcl}
\seme{c}{\env} & = & $c$ \\
\seme{a}{(A, G, T) :: \env} & = & T.a  \qquad\hspace*{0.5em}\mbox{if~} a \in A\\
\seme{a}{(A, G, T) :: \env} & = & \seme{a}{\env} \qquad\mbox{if~} a \notin A\\
\seme{\fn(\overline{e})}{\env} & = & \fn(\overline{\seme{e}{\env}}) \\
\seme{\ag(e)}{\env} & = & \ag\left(\overline{\seme{e}{((A,G,[t]) :: \env')}}\right)_{t\in T} \\
&&\mbox{if~} \findevalenv{\env}{e} = (A,G,T) :: \env'
\end{array}
\]
}
\end{minipage}
\qquad\quad
\begin{minipage}[c]{.4\linewidth}
{\footnotesize
\begin{tabular}[c]{@{\hspace{4mm}}c@{\hspace{4mm}}c@{\hspace{4mm}}c@{\hspace{4mm}}}
\infer{\findevalenv{\env}{c}  =  \env}{c \in \values} &
\multicolumn{2}{c}{\infer{\findevalenv{[]}{e}  =  {\tt undefined}}{e\notin\values}} \\[1.5ex]
\multicolumn{3}{c}{\infer{\findevalenv{((A,G,T) :: \env)}{e}  = \env'}{e\notin\values & \findevalenv{\env}{e} = \env' & \env' \neq {\tt undefined}}}\\[1.5ex]
\multicolumn{3}{c}{\infer{\findevalenv{((A,G,T) :: \env)}{e}  = (A,G,T) :: \env}{\findevalenv{\env}{e} = {\tt undefined} & {\isbuiltupon{(A \cup \bigcup_{(A',G,T)\in \env} G)}{e}}}}\\[1.5ex]
\infer{\isbuiltupon{G}{c}}{c \in \values} &
\infer{\isbuiltupon{G}{e}}{e \in G} &
\infer{\isbuiltupon{G}{\fn(\overline e)}}{\bigwedge_{\overline{e}}\isbuiltupon{G}{e}}
\end{tabular}
}
\end{minipage}

 \caption{Semantics of \sqlalg expressions.}\label{fig:algsem}
\end{figure*}

\subsection{\nrae}
\label{sec:semantics-app:nrae}

The semantics is defined \Cref{fig:nra-semantics}.
A query $q$ evaluated in a local environment $\envnraenv$ against input data $d$ produces a value~$d'$ (${\envnraenv \vdash q \rapp d \Downarrowa d'}$).
The environment~$\envnraenv$ can be any \nrae data~(\eg a record or a collection).
We refer the reader to~\citet{DBLP:conf/sigmod/AuerbachHMSS17} for a detailed explanation of the semantics.

\begin{figure*}[!h]
    \begin{minipage}[l]{\textwidth}
{\footnotesize
  \begin{gather*}
    \infer[Constant]
    {\envnraenv \vdash d_0 \qapp\ d \Downarrowa d_0}
    {}
    \qquad
    \infer[ID]
    {\envnraenv \vdash \qID \qapp\ d \Downarrowa d}
    {}
    \qquad
    \infer[Comp]
    {\envnraenv \vdash q_2 \circ q_1 \qapp\ d_0 \Downarrowa d_2}
    {\envnraenv \vdash q_1 \qapp\ d_0 \Downarrowa d_1
     \quad
     \envnraenv \vdash q_2 \qapp\ d_1 \Downarrowa d_2}
    \\[\jot]
    \infer[Unary]
    {\envnraenv \vdash \opunop\, q \qapp d \Downarrowa d_1}
    {\envnraenv \vdash q \qapp d \Downarrowa d_0
     \quad
     \opunop\, d_0 = d_1}
    \qquad
    \infer[Binary]
    {\envnraenv \vdash q_1\opbinop q_2 \qapp d \Downarrowa d_3}
    {\envnraenv \vdash q_1 \qapp d \Downarrowa d_1
     \quad
     \envnraenv \vdash q_2 \qapp d \Downarrowa d_2
     \quad
     \envnraenv \vdash d_1\opbinop d_2 = d_3}
    \\[\jot]
    \infer[Map~\emptyset]
    {\envnraenv \vdash \qmap{q_2}{q_1} \qapp d \Downarrowa \emptyset}
    {\envnraenv \vdash q_1 \qapp d \Downarrowa \emptyset}
    \qquad
    \infer[Map]
    {\envnraenv \vdash \qmap{q_2}{q_1} \qapp d \Downarrowa [d_2]\cup s_2}
    {\envnraenv \vdash q_1 \qapp d \Downarrowa [d_1]\cup s_1
     \quad
     \envnraenv \vdash q_2 \qapp d_1 \Downarrowa d_2
     \quad
     \envnraenv \vdash \qmap{q_2}{s_1} \qapp d \Downarrowa s_2}
    \\[\jot]
    \infer[Sel_{\mbox{T}}]
    {\envnraenv \vdash \qselect{q_2}{q_1} \qapp d \Downarrowa [d_1]\cup s_2}
    {\envnraenv \vdash q_1 \qapp d \Downarrowa [d_1]\cup s_1
     \quad
     \envnraenv \vdash q_2 \qapp d_1 \Downarrowa \true
     \quad
     \envnraenv \vdash \qselect{q_2}{s_1} \qapp d \Downarrowa s_2}
    \\[\jot]
    \infer[Sel_{\mbox{F}}]
      {\envnraenv \vdash \qselect{q_2}{q_1} \qapp d \Downarrowa s_2}
      {\envnraenv \vdash q_1 \qapp d \Downarrowa [d_1]\cup s_1
       \quad
       \envnraenv \vdash q_2 \qapp d_1 \Downarrowa \false
       \quad
       \envnraenv \vdash \qselect{q_2}{s_1} \qapp d \Downarrowa s_2}
    \qquad
    \infer[Sel_\emptyset]
      {\envnraenv \vdash q_1 \qapp d \Downarrowa \emptyset}
      {\envnraenv \vdash \qselect{q_2}{q_1} \qapp d \Downarrowa \emptyset}
    \\[\jot]
    \infer[Prod^{l}_{\emptyset}]
      {\envnraenv \vdash q_1\times q_2 \qapp d \Downarrowa \emptyset}
      {\envnraenv \vdash q_1 \qapp d \Downarrowa \emptyset}
    \qquad
    \infer[Prod^{r}_{\emptyset}]
      {\envnraenv \vdash q_1\times q_2 \qapp d \Downarrowa \emptyset}
      {\envnraenv \vdash q_2 \qapp d\Downarrowa \emptyset}
    \\[\jot]
    \infer[Prod]
      {\envnraenv \vdash q_1\times q_2 \qapp d \Downarrowa [d_1 \opconcat d_2]\cup s_3\cup s_4}
      {\envnraenv \vdash q_1 \!\qapp\! d\!\Downarrowa\! [d_1]\!\cup\! s_1
       \quad
       \envnraenv \vdash q_2 \!\qapp\! d\!\Downarrowa\! [d_2]\!\cup\! s_2
       \quad
       \envnraenv \vdash [d_1]\!\times\! s_2 \!\qapp\! d\!\Downarrowa\! s_3
       \quad
       \envnraenv \vdash s_1\!\times\! \left([d_2]\!\cup\! s_2\right) \!\qapp\! d\! \Downarrowa\! s_4}
    \\[\jot]
    \infer[Env]
      {\envnraenv \vdash \qENV \qapp\ d \Downarrowa \envnraenv}
      {}
    \qquad
    \infer[\mbox{Comp$^e$}]
      {\envnraenv_1 \vdash q_2 \circ^e q_1 \qapp\ d_1 \Downarrowa d_2}
      {\envnraenv_1 \vdash q_1 \qapp\ d_1 \Downarrowa \envnraenv_2
       \quad
       \envnraenv_2 \vdash q_2 \qapp\ d_1 \Downarrowa d_2}
    \\[\jot]
    \infer[\mbox{Map$^e_\emptyset$}]
      {\emptyset \vdash \qmapenv{q_2} \qapp d \Downarrowa \emptyset}
      {}
    \qquad
    \infer[\mbox{Map$^e$}]
      {[d_1]\cup s_1 \vdash \qmapenv{q_2} \qapp d \Downarrowa [d_2]\cup s_2}
      {d_1 \vdash q_2 \qapp d \Downarrowa d_2
       \quad
       s_1 \vdash \qmapenv{q_2} \qapp d \Downarrowa s_2}
    \\[\jot]
    \infer[\mbox{Default$_{\lnot\emptyset}$}]
      {\envnraenv \vdash q_1~\qor~q_2\qapp d \Downarrowa d_1}
      {\envnraenv \vdash q_1\qapp d \Downarrowa d_1
       \quad
       d_1 \neq \emptyset}
    \qquad
    \infer[\mbox{Default$_{\emptyset}$}]
      {\envnraenv \vdash q_1~\qor~q_2\qapp d \Downarrowa d_2}
      {\envnraenv \vdash q_1\qapp d \Downarrowa \emptyset
       \quad
       \envnraenv \vdash q_2\qapp d \Downarrowa d_2}
    \\[\jot]
    \infer[\mbox{Either$_{\dleft{}}$}]
      {\envnraenv \vdash \qeither{q_1}{q_2}\qapp {\dleft{}}\ d \Downarrowa d_1}
      {\envnraenv \vdash q_1\qapp d \Downarrowa d_1}
    \qquad
    \infer[\mbox{Either$_{\dright{}}$}]
      {\envnraenv \vdash \qeither{q_1}{q_2}\qapp {\dright{}}\ d \Downarrowa d_2}
      {\envnraenv \vdash q_2\qapp d \Downarrowa d_2}
  \end{gather*}
}
\end{minipage}

   \caption{\NRAEnv\ Semantics.}
  \label{fig:nra-semantics}
\end{figure*}

\subsection{\imp}
\label{sec:semantics-app:imp}

\begin{figure}[t]
  
  \centering
  \smaller
  $$
  \begin{array}[t]{lcl}
    \impsem{c}{\envimp} & = & c
    \\[0.2em]
    \impsem{x}{\envimp} & = & \envimp(x)
    \\[0.2em]
    \impsem{\mathit{op}(e)}{\envimp} & = & \impdatasem{\mathit{op}(\impsem{e}{\envimp})}
    \\[0.2em]
    \impsem{\mathit{f}(e)}{\envimp} & = & \impdatasem{\mathit{f}(\impsem{e}{\envimp})}
    \\[0.5em]

    \impsem{\impassign{x}{e}}{\envimp} & = &
       \envimp[x \leftarrow \impsem{e}{\envimp}]
    \\[0.2em]

    \impsem{\impif{e}{s_1}{s_2}}{\envimp} & = &
\algoif{\tobool{\impsem{e}{\envimp}}}{\impsem{s_1}{\envimp}}{\impsem{s_2}{\envimp}}
\\[0.2em]

    \impsem{\impforrange{x}{e}{s}}{\envimp} & = &
\algofold{(\lambda(\envimp, c). \impsem{s}{\envimp[x \leftarrow c]})}{\envimp}{(\tolist{\impsem{e}{\envimp}})}
\\[0.2em]

    \impsem{\impblock{\mathit{decls}\ \mathit{stmts}}}{\envimp} & = &
       \impsem{\mathit{stmts}}{\impsem{\mathit{decls}}{\envimp}}
         \backslash \mathit{Dom}(\mathit{decls})
    \\[0.2em]
    \impsem{s\ \mathit{stmts}}{\envimp} & = & \impsem{\mathit{stmts}}{\impsem{s}{\envimp}}
    \\[0.2em]
    \impsem{\epsilon}{\envimp} & = & \envimp
  \end{array}
  $$

   \caption{Semantics of Imp. $\tobool{c}$ and $\tolist{c}$ interpret the data value~$c$ as a Boolean or a list.}
  \label{fig:imp-sem}
\end{figure}

The semantics of \imp is defined \Cref{fig:imp-sem}. $\impsem{e}{\envimp}$ evaluates in an environment $\envimp$ an expression $e$ into a value~$c$ and $\impsem{s}{\envimp}$ evaluates statement~$s$ into a new environment~$\envimp'$.
The environment~$\envimp$ is a stack that maps from variables to values: $\envimp[x \leftarrow v]$ adds the binding of~$x$ to the value~$v$ and if~$x$ is already in~$\envimp$ it simply hides it; $\envimp \backslash \{x\}$ removes the latest binding of~$x$.

\newpage

\section{From \sqlalg to \nrae}
\label{sec:sqlatonraeapp}
This section details the full equations in the translation of \sqlalg
queries and formulas to \nrae.

\begin{figure*}[!h]
  {\small
      \[
        \setstretch{1.3}
        \begin{array}{lcl}
                    \trq{\aenv}{\mathit{tbl}} & =
          & \trtab{\mathit{tbl}}  \\
          \trq{\aenv}{Q_1~~\mathtt{union}~~Q_2} & =
          & \trq{\aenv}{Q_1}~~{ \cup}~~\trq{\aenv}{Q_2} \\
          \trq{\aenv}{Q_1~~\mathtt{intersect}~~Q_2} & =
          & \trq{\aenv}{Q_1}~~{ \cap}~~\trq{\aenv}{Q_2}\\
          \trq{\aenv}{Q_1~~\mathtt{except}~~Q_2} & =
          & \trq{\aenv}{Q_1}~~\mathtt{\setminus}~~\trq{\aenv}{Q_2} \\
          \trq{\aenv}{Q_1~~{\Large \bowtie}~~Q_2} & =
          & \trq{\aenv}{Q_1} \times \trq{\aenv}{Q_2}\\
          \trq{\aenv}{{\Large \sigma}_{f}(Q) } & =
          &  \qselect{\eapp{\pushd}{\trf{(\mathit{sort}~{Q},\{\})::\aenv}{f}}}
            {\trq{\aenv}{Q}}\\
          \trq{\aenv}{{\Large \pi}_{(\overline{e_n \mathtt{\,as\,} \attribute_n})} }(Q) & =
          & \qmap{\eapp{\pushd}
            {\trsel{(\mathit{sort}~{Q},\{\})::\aenv}{\overline{e_n \mathtt{\,as\,} \attribute_n}}}}
            {\trq{\aenv}{Q}} \\
          \trq{\aenv}{{\large \gamma}_{(\overline{e_n \mathtt{\,as\,} \attribute_n},\overline{b_k},f)}(Q) } & =
          & \\
          \multicolumn{3}{c}
          {\qquad\qmap{\eapp{\pushq}
          {\trsel{(\mathit{sort}~{Q},\overline{b_k})::\aenv}{\overline{e_n \mathtt{\,as\,} \attribute_n}}}}
          {\qselect{\eapp{\pushq}
          {\trf{(\mathit{sort}~{Q},\overline{b_k})::\aenv}{f}}}
          {\qmap{\tdot{\qID}{g}}{\qgroupby{g}{\overline{b_k}}{\trq{\aenv}{Q}}}}}}\\
          \multicolumn{3}{l}{\hspace*{1.6cm}\mbox{where}~g~
          \mbox{is a fresh label \wrt}~\overline{b_k}}
\end{array}
      \]
    }

   \caption{Compilation of \sqlalg queries to \nrae.}
\end{figure*}

\begin{figure}[!h]
  {\small
      \[
        \setstretch{1.3}
        \begin{array}[t]{lcl}
          \trf{\aenv}{f_{1}~~\mathtt{and}~~f_{2}} & =
          & (\trf{\aenv}{f_{1}})~~\wedge_B~~(\trf{\aenv}{f_{2}})  \\
          \trf{\aenv}{f_{1}~~\mathtt{or}~~f_{2}} & =
          & (\trf{\aenv}{f_{1}})~~\vee_B~~(\trf{\aenv}{f_{2}}) \\
          \trf{\aenv}{ \mathtt{not}~~f} & =
          & \neg_B~~(\trf{\aenv}{f})  \\
          \trf{\aenv}{p~ (\overline{e})} & =
          &  \trp{p}{(\overline{\tre{\aenv}{e}})}  \\
          \trf{\aenv}{ \mathtt{true}} & =
          & \mathtt{true_B} \\
        \end{array}
        \quad
        \begin{array}[t]{lcl}
          \trf{\aenv}{p~~(e,\mathtt{all} ~~\algquery)} & =
          & \mathtt{all}_{B}~~p~~(\tre{\aenv}{e})~
            ~(\trq{\aenv}{\algquery})\\
          \trf{\aenv}{p~~(e,\mathtt{any} ~~\algquery)} & =
          & \mathtt{any}_B~~p~~(\tre{\aenv}{e})~
            ~(\trq{\aenv}{\algquery})\\
          \trf{\aenv}{ls~~ \mathtt{in}~~\algquery)} & =
          & (\trsel{\aenv}{ls})~~\mathtt{in}_B~
            ~(\trq{\aenv}{\algquery})\\
          \trf{\aenv}{\mathtt{exists} ~~\algquery} & =
          & \mathtt{count} ~
            ~(\trq{\aenv}{\algquery}) \mathtt{>} 0\\
        \end{array}
      \]
    }

   \caption{Compilation of \sqlalg formulas to \nrae.}\label{fig:ftrans}
\end{figure}

\newpage

\section{Benchmarks}
\label{sec:evaluation-app}

\subsection{\null Queries}

Instance:
\begin{lstlisting}[language=SQLL]
{ "R": [ { "A": null },
         { "A": 1.0 } ],
  "S": [ { "A": null } ],
  "T": [ { "A": null },
         { "A": null },
         { "A": 1.0 } ] }
\end{lstlisting}

\noindent
Queries:
\begin{lstlisting}[language=SQLL, mathescape]
create table R (A double precision);
create table S (A double precision);
create table T (A double precision);

select R.A from R where R.A not in (select S.A from S);
-- Expected: []

select R.A from R where not exists (select * from S where S.A = R.A);
-- Expected: [{"A":1},{"A":null}]

select R.A from R except select S.A from S;
-- Expected: [{"A":1}]

select T.A, count( * ) as c from T group by T.A;
-- Expected: [{"A":null, "c":2},{"A":1,"c":1}]
\end{lstlisting}

\subsection{Correlated Queries}

\noindent
Instance:
\begin{lstlisting}[language=SQLL]
{ "t1": [ { "a1": 1.0, "b1": 1.0 },
          { "a1": 1.0, "b1": 2.0 },
          { "a1": 1.0, "b1": 3.0 },
          { "a1": 1.0, "b1": 4.0 },
          { "a1": 1.0, "b1": 5.0 },
          { "a1": 1.0, "b1": 6.0 },
          { "a1": 1.0, "b1": 7.0 },
          { "a1": 1.0, "b1": 8.0 },
          { "a1": 1.0, "b1": 9.0 },
          { "a1": 1.0, "b1": 10.0 },
          { "a1": 2.0, "b1": 1.0 },
          { "a1": 2.0, "b1": 2.0 },
          { "a1": 2.0, "b1": 3.0 },
          { "a1": 2.0, "b1": 4.0 },
          { "a1": 2.0, "b1": 5.0 },
          { "a1": 2.0, "b1": 6.0 },
          { "a1": 2.0, "b1": 7.0 },
          { "a1": 2.0, "b1": 8.0 },
          { "a1": 2.0, "b1": 9.0 },
          { "a1": 2.0, "b1": 10.0 },
          { "a1": 3.0, "b1": 1.0 },
          { "a1": 3.0, "b1": 2.0 },
          { "a1": 3.0, "b1": 3.0 },
          { "a1": 3.0, "b1": 4.0 },
          { "a1": 3.0, "b1": 5.0 },
          { "a1": 4.0, "b1": 6.0 },
          { "a1": 4.0, "b1": 7.0 },
          { "a1": 4.0, "b1": 8.0 },
          { "a1": 4.0, "b1": 9.0 },
          { "a1": 4.0, "b1": 10.0 } ],
  "t2": [ { "a2": 7.0, "b2": 7.0 },
          { "a2": 7.0, "b2": 7.0 } ] }
\end{lstlisting}

\noindent
Queries:
\begin{lstlisting}[language=SQLL, mathescape]
create table t1 (a1 double precision, b1 double precision);
create table t2 (a2 double precision, b2 double precision);

select a1, max(b1) from t1 group by a1;
-- Expected: (a1=1,max=10); (a1=2,max=10); (a1=3,max=5); (a1=4,max=10)

select a1 from t1 group by a1 having exists (select a2 from t2 group by a2 having sum(1.0+0.0*a1) = 10.0);
-- Expected: (a1=1); (a1=2)

select a1 from t1 group by a1 having exists (select a2 from t2 group by a2 having sum(1.0+0.0*a2) = 10.0);
-- Expected: empty

select a1 from t1 group by a1 having exists (select a2 from t2 group by a2 having sum(1.0+0.0*a2) = 2.0);
-- Expected: (a1=1); (a1=2); (a1=3); (a1=4)

select a1 from t1 group by a1 having exists (select a2 from t2 group by a2 having sum(1.0) = 2.0);
-- Expected: (a1=1); (a1=2); (a1=3); (a1=4)

select a1 from t1 group by a1 having exists (select a2 from t2 group by a2 having sum(1.0) = 10.0);
-- Expected: empty

select a1 from t1 group by a1 having exists (select a2 from t2 group by a2 having sum(1.0+0.0*a1)+sum(1.+0.0*a2) = 12.0);
-- Expected: (a1=1); (a1=2)

select a1 from t1 group by a1 having exists (select a2 from t2 group by a2 having sum(1.0+0.0*a1+0.0*a2) =2.0);
-- Expected: (a1=1); (a1=2); (a1=3); (a1=4)

select a1 from t1 group by a1 having exists (select a2 from t2 group by a2 having sum(1.0+0.0*a1+0.0*a2) = 3.0);
-- Expected: empty

select a1 from t1 group by a1 having exists (select a2 from t2 group by a2 having sum(1.0+0.0*a1+0.0*b2) = 2.0);
-- Expected: (a1=1); (a1=2); (a1=3); (a1=4)

select a1 from t1 group by a1 having exists (select a2 from t2 group by a2 having sum(1.0+0.0*a1+0.0*b2) = 3.0);
-- Expected: empty  
\end{lstlisting}

\subsection{Performance Evaluation}
\label{sec:related:evaluation:performance}

In addition to semantic correctness, we also kept performance in mind during the design of \dbcert. In order to ensure reasonable runtime behavior for standard SQL queries:
\begin{itemize}
\item The runtime is equipped with an efficient \lstjs+push+ operation
  on persistent arrays (\Cref{sec:implementation:runtime});
\item Several optimizations are implemented and proved correct, complementing those pre-existing in 
Q*cert~\cite{DBLP:conf/sigmod/AuerbachHMSS17a}.
\end{itemize}
Those optimizations present as improvements in the \sqlalg to \nrae
translation~(\Cref{sec:sqlatonrae:translation:queries}) and as
rewrites on later intermediate representations. Most of our existing
optimization efforts focus on eliminating inefficiencies introduced by
translation.  Notably, operations on \null introduce complex
expressions that can often be simplified (e.g.,
${\qeither{q_1}{q_2}} \circ {\dleft{q}} = q_1 \circ q$). This is part
of our proof strategy: rather than trying to prove a complex but more
efficient translation, we implement and prove correct a naive, less
efficient translation, then verify and apply individual rewrites.

We sketched the performance on very preliminary experiments, using the
version of our compiler with floating point values. We used a simple
table, named \Sql!employees!, with two fields, \Sql!name! (of type
\Sql!text!) and \Sql!age! (of type
\Sql!double precision!), populated with 58,800 entries in the JSON
format. We run various queries, with and without aggregates and grouping
operations; examples of queries are:
\begin{lstlisting}[language=SQLL]
select avg(age) from employees where age > 32.0;
select age, count(*) from employees group by age;
\end{lstlisting}
These experiments shows that \dbcert runs about two times slower than
\alasql. Further optimizations and performance improvements are future
work.

\fi{}

\end{document}